\documentclass{aa}  
\usepackage{graphicx}
\usepackage{txfonts}
\usepackage{hyperref}

\hypersetup{colorlinks,linkcolor={blue},citecolor={blue},urlcolor={blue}}  
\usepackage{diagbox}
\usepackage{threeparttable}
\usepackage{lscape}
\usepackage{longtable}
 
\usepackage{tikz,hyperref}
\usepackage{academicons}
\definecolor{lime}{HTML}{A6CE39}
\DeclareRobustCommand{\orcidicon}{%
        \begin{tikzpicture}
        \draw[lime, fill=lime] (0,0) 
        circle [radius=0.16] 
        node[white] {{\fontfamily{qag}\selectfont \tiny ID}};
        \draw[white, fill=white] (-0.0625,0.095) 
        circle [radius=0.007];
        \end{tikzpicture}
        \hspace{-2mm}
}

\foreach \x in {A, ..., Z}{%
        \expandafter\xdef\csname orcid\x\endcsname{\noexpand\href{https://orcid.org/\csname orcidauthor\x\endcsname}{\noexpand\orcidicon}}
}
\newcommand{\orcid}[1]{\href{https://orcid.org/#1}{\textcolor[HTML]{A6CE39}{\orcidicon}}}

\newcommand{\zenodo}{https://zenodo.org/records/14510557?token=eyJhbGciOiJIUzUxMiJ9.eyJpZCI6ImExMDZmNmViLTYzMjctNGU5Ni04NmJiLTMwYWJkYWJkZTgxMyIsImRhdGEiOnt9LCJyYW5kb20iOiJjYWZjZDRlMzNlZTI3NDdmYzc3ZTAyNmI1NzI0OTk4OSJ9.ImVsR86xkrgOvRhVAXRSZWyfRsIf0wUVnaphmN5mbnH_aBA28cm3Rf5dqmFxCIHIfj-HdxGKuV093t6Mv8-41A}

\newcommand{\xmm}{{\it XMM-Newton}\xspace}

\newcommand{\swift}{{\it Swift}\xspace}

\newcommand{\eqb}{\begin{eqnarray}}
\newcommand{\eqe}{\end{eqnarray}}
\defcitealias{Haberl2016}{HS16}

\graphicspath{{./}{finalplots/}}
\begin{document} 
\title{Two decades of optical variability of Small Magellanic Cloud high-mass X-ray binaries} 
   \titlerunning{Optical variability of SMC HMXBs}
   \author{H. Treiber\orcid{0000-0003-0660-9776}\inst{1,2,3}\thanks{\email{lena.treiber@princeton.edu}},
          G. Vasilopoulos\orcid{0000-0003-3902-3915}\inst{4,5}\thanks{\email{gevas@phys.uoa.gr}},
          C. D. Bailyn\inst{1},
          F. Haberl\orcid{0000-0002-0107-5237}\inst{6},
          A. Udalski\orcid{0000-0001-5207-5619}\inst{7} 
          }
    \authorrunning{H. Treiber et al.}
   \institute{Department of Astronomy, Yale University, PO Box 208101, New Haven, CT 06520-8101, USA
   \and Department of Physics and Astronomy, Amherst College, C025 New Science Center, 25 East Dr., Amherst, MA 01002-5000, USA
   \and
   Department of Astrophysical Sciences, Princeton University, 4 Ivy Lane, Princeton, NJ 08544, USA
   \and 
   Department of Physics, National and Kapodistrian University of Athens, University Campus Zografos, GR 15784, Athens, Greece
    \and
    Institute of Accelerating Systems \& Applications, University Campus Zografos, Athens, Greece
    \and
    Max-Planck-Institut f{\"u}r extraterrestrische Physik, Gie{\ss}enbachstra{\ss}e 1, D-85748 Garching, Germany
        \and
    Astronomical Observatory, University of Warsaw, Al. Ujazdowskie 4, 00-478, Warszawa, Poland
    \\
    }
   \date{Received 15 June 2024; accepted 10 December 2024}

  \abstract
   {}
   {We present an analysis of the long-term optical/IR behavior of 111 high-mass X-ray binaries (HMXBs) in the Small Magellanic Cloud based on data from the OGLE collaboration. }
   {Most systems exhibit variability on a range of time scales. This variability regulates the mass transfer to the compact object, while the compact object can, in turn, affect the donor star's behavior. To better understand this complex interaction and the resulting X-ray properties in these systems, we define a new taxonomy for the observed super-orbital variability.}
   {This taxonomy connects to the color changes, orbital periods, and X-ray behavior of the sources. In most cases, these properties can be explained by differences between the flux of the disk around the Be star and the flux from the star itself. We also refine and present new potential orbital periods and sub-orbital variability in the sources.}
   {}    
   \maketitle
%
\section{Introduction}
High-mass X-ray binaries (HMXBs) consist of a compact object, either a neutron star (NS) or a black hole, in orbit around an O- or B-type star. These sources are likely precursors of double compact binaries, and some may evolve into gravitational wave sources \citep[e.g.,][]{Gies2019}.
Accretion from the early-type star onto the compact object results in powerful X-ray emission, making HMXBs some of the brightest X-ray sources in the sky.

In one common subclass of HMXB, the high-mass donor is a Be star, and the accreting source is an NS. These systems are referred to as BeXRBs \citep[for a review see][]{Reig2011}.
The Be stars rotate rapidly and host ``decretion" disks of ejected ionized gas, which are observable through emission lines and infrared excess \citep[][]{Harmanec1982,Rivinius2013}. The disk often dominates the observed optical and IR variability from these systems. On the other hand, the X-ray emission arises from accretion from the Be disk onto hotspots on the NS magnetic poles. 
The brightest X-ray emission from these sources occurs during type I and type II outbursts. Type I X-ray outbursts ($\mathrm{L_{X} \sim 10^{36}-10^{37} erg\,s^{-1}}$) are typically periodic, coinciding with periastron passage. A few observed BeXRBs have low eccentricity orbits ($e\lesssim0.3$) and do not exhibit this transient form of accretion \citep[e.g.,][]{Pfahl2002}, although some $e\lesssim0.3$ systems have significant outbursts \citep[e.g.,][]{Haberl2016}. Type II outbursts ($\mathrm{L_{X} \gtrsim 10^{37} erg\,s^{-1}}$) can last longer than an orbit and likely involve the transient formation of an accretion disk around the NS. 
On shorter time scales, pulsations from the Be star can be observed in the optical/IR \citep[e.g.,][]{Rivinius2003}, and the NS spin period is often observable in the X-ray as the hotspots sweep in and out of view \citep{Davidson1973}.

Long-term optical/IR monitoring of these systems is essential in the pursuit of a complete understanding of BeXRBs, including the evolutionary pathways that lead to double degenerate systems. The variability of isolated Be stars is itself poorly understood \citep[e.g.,][]{Mennickent2002,Kourniotis2014,Labadie-Bartz2017}, and the interactions of the NS and Be star disk further complicate matters, as the NS can truncate or entirely deplete the disk \citep{Reig1997,Negueruela2001,Okazaki2001,Okazaki2002}.
Periodic modulation may be related to the orbital period or pulsations of the star, while longer-term variability on the order of thousands of days is a feature of Be stars in general \citep[e.g.,][]{deWit2006,Kourniotis2014}, but it can also be a result of disk precession and truncation effects driven by the NS \citep{Martin2011}. Past studies have focused on Be star non-radial pulsations \citep[e.g.,][]{Schmidtke2013}, potential orbital periodicities \citep[e.g.,][]{Bird2012}, super-orbital modulations \citep[e.g.,][]{Rajoelimanana2011}, and color-magnitude properties \citep[e.g.,][]{deWit2006,Rajoelimanana2011}.

As nearby sites of star formation, the Small and Large Magellanic Clouds (SMC and LMC, respectively) are prime targets for the identification and characterization of HMXBs. 
Simplifying the analysis of these systems are the well-constrained distances of $\sim$50 kpc \citep[LMC,][]{Pietrzynski2019} and $\sim$62 kpc \citep[SMC,][]{Graczyk2020} as well as a lack of contamination from the Galactic plane. In addition, stellar population studies provide constraints on the metallicity and age of the underlying stellar populations \citep[e.g.,][]{Massey1995,Milone2009}.

A particularly useful dataset for studying variability in the LMC and SMC is provided by the Optical Gravitational Lensing Experiment \citep[OGLE;][]{1992AcA....42..253U, 2008AcA....58...69U,Udalski2015}, which has monitored the Magellanic Clouds since 1992. 
Starting with OGLE-II in 1997, images have been taken at Las Campanas Observatory in V- and I-bands with the 1.3\,m Warsaw telescope. The subsequent phases OGLE-III and OGLE-IV involved improvements to the charge-coupled devices and an increased field of view. The OGLE database\footnote{OGLE XROM portal: \url{https://ogle.astrouw.edu.pl/ogle4/xrom/xrom.html}} provides the light curves of most known X-ray binaries in the Magellanic Clouds and \citet{2008AcA....58...69U} describes the data reduction.

In this paper, we present an analysis of OGLE's monitoring of 111 HMXBs in the SMC, 63 of which have measured NS spin periods and most of which are BeXRBs.
We used data from OGLE II, III, and IV to study the orbital and super-orbital variability and the colors of these systems. 
We define a new classification for the super-orbital variability of these sources, which we interpret as an expression of the changing flux ratio between the Be star and its disk. 
We also search the detrended light curves for evidence of orbital and non-radial pulsation periods, and we explore the relationship between the optical/IR behavior, the orbital period, and the spin period of the NS.

In Sect. \ref{sec:sample}, we explain the construction of our sample.
In Sect. \ref{sec:tax}, we present five types of super-orbital variability based on two parameters extracted from the OGLE data: the relationship between the median and brightness extrema and the relationship between the stochastic and coherent variability. The sample's color-magnitude properties are presented in Sect. \ref{sec:color}, and we outline our search for orbital periodicities in Sect. \ref{sec:period}. In Sect. \ref{sec:interesting}, we highlight a few sources that have particularly interesting properties.
We discuss the relationship between the phenomenology we observe and the properties of the Be star disk in Sect. \ref{sec:discussion}. Lastly, in Sect. \ref{sec:conclusion}, we summarize our findings and provide recommendations for future work.

\section{OGLE sample of HMXBs in the Small Magellanic Cloud}
\label{sec:sample}
To construct our sample, we started with a recent catalog of HMXBs in the SMC \citep[i.e.,][hereafter HS16]{Haberl2016}. The original catalog was composed of 148 sources, of which 121 were relatively high-confidence HMXBs. 
We included sources with both measured spin periods and proposed optical counterparts, but based the remainder of our sample on the \citetalias{Haberl2016} highly confident HMXBs sample.
However, because of additions and corrections from the recent literature, we clarify a few selection criteria and updates to the catalog.
First, we only included sources with low uncertainty in X-ray position (i.e., $<$10$^{\prime\prime}$). Thus, we excluded pulsars only detected by RXTE, and without any recent X-ray detection with imaging instruments. 

Since \citetalias{Haberl2016}, monitoring surveys \citep[e.g., S-CUBED][]{S-CUBED} of the SMC have revealed a number of outbursts that were followed-up by other X-ray telescopes (e.g., \xmm, \swift, NICER) enabling identification or better characterization of the sources
SXP\,2.16 \citep[][]{Vasilopoulos2017}, SXP\,4.78 \citep{Strohmayer2018,Schmidtke2022}, SXP\,15.6 \citep{Vasilopoulos2017,Vasilopoulos2022}, SXP\,305 \citep{Lazzarini2019}, SXP 164 \citep{Carpano2022}, SXP\,182 \citep[][]{Coe2023,Gaudin2024}, and SXP\,342B \citep[][]{Maitra2023}.

In recent years, OGLE IV has included high-cadence (15 minute) data for many sources in the sample. This new view of the population is incredibly informative, but its analysis is beyond the scope of the current study. We exclude these data from our periodicity searches and bin the data to a two-day cadence for the other aspects of our analysis.

Our sample is listed in Tables \ref{t.xray} and \ref{t.optical}.  Table \ref{t.xray} provides prior information about our sources: the \citetalias{Haberl2016} identification number, the X-ray
coordinates and error, the coordinates and ID of the OGLE counterparts, and X-ray spin and X-ray orbital periods when known.  Table \ref{t.optical} reports the results of our detailed investigation of the OGLE light curves. We note that for 21 sources with unphysical jumps between OGLE III and IV, we calibrated OGLE IV measurements by setting their median equal to the overall prior median.
Using the convention introduced by \citet{Coe2005}, we also refer to the sources with measured spin periods by ``SXP" (i.e., SMC X-ray pulsar) followed by the pulse period to three significant figures.

\section{Super-orbital taxonomy of optical light curves}
\label{sec:tax}

\begin{table*}
\begin{tiny}
\caption{Definition of source taxonomy used in this paper.}\label{t.type_definitions}  
\begin{tabular}{llll}
\textbf{Type \#} & \textbf{Base Number} & \textbf{Stochastic Variability Metric} & \textbf{Qualitative Description} \\
\hline
\hline\noalign{\smallskip} 
1&<$-$0.55&<0.35&faint base, similarly shaped peaks\\
2&>0.15&<0.35&well-defined brightness maximum, dips\\
3&>$-$0.55 and <0.15&<0.35&no faint base or consistent, well-defined brightness maximum, but comparably variable\\
4&>$-$0.55 and <0.15&>0.40 and <0.75&coherent super-orbital variability present but does not dominate the overall light curve variability\\
5 & & >0.75&only short-term variability      
\end{tabular}
\end{tiny}
\end{table*}

\longtab{
\begin{tiny}
\begin{longtable}{llllllllll}
\caption{\label{t.xray} Identifying information for HMXBs in our sample.} \\
\textbf{Source \#} & \textbf{H\&S (2016) \#} & \textbf{X-ray RA} & \textbf{X-ray Dec} & \textbf{Pos Err} & \textbf{OGLE RA} & \textbf{OGLE Dec} & \textbf{$\mathrm{P_{spin}}$ (s)} & \textbf{X-ray $\mathrm{P_{orb}}$ (d)} & \textbf{OGLE ID} \\
\hline
\hline\noalign{\smallskip} 
\endfirsthead
\caption{continued.}\\
\textbf{Source \#} & \textbf{H\&S (2016) \#} & \textbf{X-ray RA} & \textbf{X-ray Dec} & \textbf{Pos Err} & \textbf{OGLE RA} & \textbf{OGLE Dec} & \textbf{$\mathrm{P_{spin}}$ (s)} & \textbf{X-ray $\mathrm{P_{orb}}$ (d)} & \textbf{OGLE ID} \\
\hline
\hline\noalign{\smallskip} 
\endhead
\hline
\endfoot
1&1&01:17:05.2&-73:26:36&0.5&01:17:05.14&-73:26:36.0&0.717&3.89&smc733.21.1392D\\ 
2&2,142&01:21:41.0&-72:57:33&4.0&01:21:40.61&-72:57:30.9&2.165&82&smc732.03.3540\\
3\textsuperscript{a}&3&00:54:33.5&-73:41:01&2&00:54:33.44&-73:41:01.3&2.37&18.38&smc720.17.50\\
4\textsuperscript{b} &4&00:59:12.5&-71:38:45&2.2&00:59:12.74&-71:38:44.9&2.763&$\dots$&smc718.01.10792\\
5\textsuperscript{c} &5&00:51:39.2&-72:17:03&1.4&00:51:38.79&-72:17:04.8&4.78&$\dots$&smc719.21.22049\\
6&6&00:57:02.3&-72:25:55&0.5&00:57:02.19&-72:25:55.4&5.05&17.13&smc719.18.378\\
7&7&01:02:53.4&-72:44:35&0.5&01:02:53.31&-72:44:35.1&6.85&21.9&smc726.29.18\\
8&9&00:52:05.7&-72:26:04&0.6&00:52:05.63&-72:26:04.1&7.78&44.92&smc719.20.144\\
9&10&00:57:58.4&-72:22:30&1.5&00:57:58.51&-72:22:29.2&7.92&$\dots$&smc719.18.465\\
10&11&00:51:53.2&-72:31:48&1.0&00:51:53.16&-72:31:48.6&8.9&28.47&smc719.12.35646\\
11&13&01:04:42.3&-72:54:04&0.6&01:04:42.30&-72:54:04.2&11.48&36.3&smc726.20.19470\\
12&14&01:57:16.0&-72:58:33&3.8&01:57:16.17&-72:58:32.5&11.58&$\dots$&mbr102.01.984\\
13&15&00:48:14.0&-73:22:04&0.5&00:48:14.04&-73:22:04.0&11.866&$\dots$&smc720.28.190\\
14&16&00:52:14.0&-73:19:18&1.0&00:52:13.99&-73:19:18.8&15.3&$\dots$&smc720.26.47\\
15&72&00:48:55.6&-73:49:46&0.6&00:48:55.36&-73:49:45.7&15.64&$\dots$&smc720.11.13342\\
16&18&00:49:11.5&-72:49:36&0.5&0:49:11.45&-72:49:37.1&18.37&17.73&smc101.8.21127 (III)\\
17&19&01:17:40.4&-73:30:51&1.3&01:17:40.16&-73:30:50.6&22.07&$\dots$&smc733.21.3978\\
18&20&00:48:14.2&-73:10:04&0.6&00:48:14.10&-73:10:03.9&25.55&$\dots$&smc720.28.40482\\
19&21&01:11:08.6&-73:16:46&0.7&01:11:08.58&-73:16:46.3&31.03&$\dots$&smc726.08.69\\
20&22&00:53:55.2&-72:26:46&0.6&00:53:55.32&-72:26:45.3&46.63&137.36&smc719.20.34\\
21&23&00:54:56.3&-72:26:47&0.6&00:54:56.18&-72:26:47.8&59.07&122.1&smc719.19.162\\
22&24&01:07:12.6&-72:35:34&0.9&01:07:12.60&-72:35:33.9&65.78&$\dots$&smc726.27.20665\\
23&25&00:49:03.3&-72:50:52&0.6&00:49:03.33&-72:50:52.5&74.67&61.6&smc719.05.41432\\
24&26&00:52:08.9&-72:38:03&0.6&00:52:08.95&-72:38:03.2&82.4&362.3&smc719.11.36482\\
25&27&00:50:57.1&-72:13:33&0.5&00:50:56.99&-72:13:34.4&91.12&87.6&smc719.21.21951\\
26&29&00:57:27.1&-73:25:19&0.8&00:57:27.06&-73:25:19.4&101.16&$\dots$&smc726.15.77\\
27&30&00:53:24.0&-72:27:16&1.0&00:53:23.86&-72:27:15.4&138.0&103.6&smc719.20.325\\
28&31&00:56:05.7&-72:21:59&0.9&00:56:05.54&-72:21:59.6&140.1&$\dots$&smc719.18.27492\\
29&$\dots$&00:45:17.91&-73:47:05.5&1.9&0:45:17.77&-73:47:05.7&146.6&$\dots$&smc720.12.13583\\ 
30&33&00:57:50.3&-72:07:57&1.0&00:57:50.38&-72:07:56.3&152.34&$\dots$&smc719.26.179\\
31&34&01:07:43.3&-71:59:54&0.6&01:07:43.42&-71:59:53.9&153.99&$\dots$&smc725.12.12425\\
32&132&01:04:29.4&-72:31:37&1.3&1:04:29.12&-72:31:37.1&164&$\dots$&smc726.28.23178\\ 
33&35&00:52:55.1&-71:58:06&0.5&00:52:55.28&-71:58:06.3&169.3&68.54&smc719.28.16997\\
34&36&00:51:51.9&-73:10:33&0.9&00:51:52.02&-73:10:34.0&172.0&70.42&smc720.26.36274\\
35&37&01:01:52.3&-72:23:33&0.5&01:01:52.29&-72:23:34.1&175.4&$\dots$&smc725.06.151\\
36&\dots&01:09:02.72&-72:37:11.7&2.1&01:09:02.25&-72:37:10.1&182&\dots&smc726.26.15515s\\
37&38&00:59:21.0&-72:23:17&0.5&00:59:21.03&-72:23:17.4&201.9&$\dots$&smc719.17.49\\
38&39&00:59:28.9&-72:37:04&0.5&00:59:28.67&-72:37:04.2&202.52&$\dots$&smc719.08.27301\\
39&40&00:50:11.1&-73:00:25&0.5&00:50:11.26&-73:00:26.0&214.03&$\dots$&smc719.04.754\\
40&41&00:47:23.3&-73:12:28&0.5&0:47:24.17&-73:12:26.6&263.0&$\dots$&smc\_sc4.116980 (II)\\
41&43&00:57:49.4&-72:02:36&0.5&00:57:49.57&-72:02:36.1&280.4&64.8&smc719.26.19885\\
42&44&00:58:12.7&-72:30:48&0.5&00:58:12.59&-72:30:48.8&291.33&151&smc719.17.2\\
43&45&00:50:48.1&-73:18:18&0.9&00:50:47.99&-73:18:18.0&292.7&$\dots$&smc720.27.303\\
44&46&01:01:02.8&-72:06:58&1.2&01:01:02.89&-72:06:59.1&304.5&$\dots$&smc725.15.85\\
45&93&00:52:15.4&-73:19:15&1.0&00:52:15.39&-73:19:15.4&305.69&$\dots$&smc720.26.531\\
46&47&00:50:44.6&-73:16:05&0.6&00:50:44.71&-73:16:05.4&323.2&116.60&smc720.27.43155\\
47&48&00:52:52.1&-72:17:15&0.6&00:52:52.23&-72:17:15.1&325.4&$\dots$&smc719.20.27652\\
48&\dots&01:24:22.9&-72:42:48.7&1.4&01:24:22.79&-72:42:48.7&341.7&\dots&smc732.10.7\\ 
49&49&00:54:03.9&-72:26:32&0.6&00:54:03.86&-72:26:32.8&341.9&$\dots$&smc719.19.166\\
50&50&01:03:13.9&-72:09:14&0.9&01:03:13.91&-72:09:14.3&345.2&$\dots$&smc725.14.251\\
51&51&01:01:20.7&-72:11:19&0.5&01:01:20.66&-72:11:19.1&455.0&$\dots$&smc725.15.42\\
52&52&00:54:55.9&-72:45:11&0.5&00:54:55.88&-72:45:10.8&499.2&265.3&smc719.10.139\\
53&53&01:02:47.5&-72:04:51&0.8&01:02:47.58&-72:04:51.4&522.5&$\dots$&smc725.14.375\\
54&54&00:57:36.2&-72:19:34&0.7&00:57:36.01&-72:19:34.1&565.0&151.8&smc719.18.27563\\
55&55&00:55:35.4&-72:29:07&0.5&00:55:35.15&-72:29:06.6&644.6&$\dots$&smc719.19.125\\
56&56&00:55:18.3&-72:38:52&0.5&00:55:18.44&-72:38:51.9&701.6&$\dots$&smc719.10.33543\\
57&57&01:05:55.4&-72:03:49&0.5&01:05:55.25&-72:03:50.6&726.0&$\dots$&smc725.12.86\\
58&58&00:49:42.1&-73:23:15&0.5&00:49:42.00&-73:23:14.6&755.5&389.9&smc720.27.17\\
59&59&00:49:29.7&-73:10:59&0.6&0:49:29.81&-73:10:58.0&894.0&$\dots$&smc100.7.58770 (III)\\
60&60&01:02:06.7&-71:41:16&0.8&01:02:06.67&-71:41:16.2&967.0&$\dots$&smc725.23.10589\\
61&61&01:27:46.0&-73:32:56&0.6&01:27:45.96&-73:32:56.4&1062.0&$\dots$&smc738.07.1966\\
62&62&01:03:37.5&-72:01:33&1.1&01:03:37.54&-72:01:33.2&1323.0&$\dots$&smc725.14.19468\\
63&63&00:54:46.2&-72:25:23&1.0&00:54:46.37&-72:25:22.8&4693.0&$\dots$&smc719.19.181\\
64&65&00:42:07.8&-73:45:03&0.7&0:42:07.87&-73:45:01.9&$\dots$&$\dots$&smc128.4.317 (III)\\
65&66&00:43:15.9&-73:24:39&1.5&0:43:15.91&-73:24:37.0&$\dots$&$\dots$&smc713.08.444\\ 
66&67&00:45:00.2&-73:42:47&1.7&0:45:00.06&-73:42:45.5&$\dots$&$\dots$&smc720.22.154\\
67&69&00:48:18.7&-73:21:00&0.6&0:48:18.66&-73:21:00.2&$\dots$&$\dots$&smc720.28.522\\ 
68&70&00:48:34.1&-73:02:31&0.7&0:48:34.11&-73:02:31.3&$\dots$&$\dots$&smc719.05.239\\ 
69&71&00:48:49.0&-73:16:25&4.4&0:48:49.31&-73:16:24.4&$\dots$&$\dots$&smc720.28.39745\\ 
70&74&00:49:13.6&-73:11:38&0.5&00:49:13.61&-73:11:37.8&\dots&\dots&smc720.28.40399\\
71&75&00:49:22.2&-73:20:06&3.4&0:49:21.75&-73:20:06.2&$\dots$&$\dots$&smc720.28.1295\\ 
72&76&00:49:30.62&-73:31:09.4&0.6&00:49:30.5&-73:31:09.2&$\dots$&$\dots$&smc720.20.4843D\\ 
73&77&00:49:41.7&-72:48:43&0.6&0:49:42.14&-72:48:43.5&$\dots$&$\dots$&smc719.13.7012\\ 
74&78&00:50:04.4&-73:14:26&1.6&0:50:04.45&-73:14:27.1&$\dots$&$\dots$&smc720.27.43471\\ 
75&79&00:50:12.2&-73:11:56&1.7&0:50:12.42&-73:11:56.4&$\dots$&$\dots$&smc720.27.43256\\ 
76&80&00:50:35.5&-73:14:01&1.1&0:50:35.60&-73:14:02.5&$\dots$&$\dots$&smc720.27.43493\\ 
77&81&00:50:36.0&-73:17:39&0.9&0:50:36.12&-73:17:39.6&$\dots$&$\dots$&smc720.27.311\\ 
78&83&00:50:47.8&-73:17:36&1.0&0:50:47.82&-73:17:36.6&$\dots$&$\dots$&smc720.27.768\\ 
79&84&00:50:57.3&-73:10:08&0.5&0:50:57.12&-73:10:08.1&$\dots$&$\dots$&smc720.27.43092\\ 
80&85&00:51:05.7&-73:13:12&1.2&0:51:05.65&-73:13:11.8&$\dots$&$\dots$&smc720.27.43232\\ 
81&86&00:51:17.0&-73:16:06&1.0&0:51:17.11&-73:16:06.9&$\dots$&$\dots$&smc720.27.43153\\ 
82&92&00:52:07.8&-72:21:26&2.0&0:52:07.43&-72:21:25.6&$\dots$&$\dots$&smc719.20.27106\\ 
83&95&00:52:37.3&-72:27:32&1.2&0:52:37.29&-72:27:32.3&$\dots$&$\dots$&smc719.20.117\\ 
84&96&00:52:45.0&-72:28:44&1.0&0:52:45.10&-72:28:43.7&$\dots$&$\dots$&smc719.20.104\\ 
85&97&00:52:52.2&-72:48:30&1.0&0:52:52.31&-72:48:30.1&$\dots$&$\dots$&smc719.11.97\\ 
86&98&00:52:59.5&-72:54:02&2.1&0:52:59.50&-72:54:03.3&$\dots$&$\dots$&smc719.03.44671\\ 
87&100&00:53:18.5&-72:16:18&1.6&0:53:18.29&-72:16:15.8&$\dots$&$\dots$&smc719.20.27695\\ 
88&101&00:53:29.2&-72:33:48&2.7&0:53:29.31&-72:33:48.3&$\dots$&$\dots$&smc719.11.36421\\
89&104&00:53:41.8&-72:53:10&0.8&0:53:42.19&-72:53:09.7&$\dots$&$\dots$&smc719.03.43645\\ 
90&105&00:53:52.5&-72:26:39&0.9&0:53:52.11&-72:26:38.9&$\dots$&$\dots$&smc719.20.36\\ 
91&106&00:54:08.7&-72:32:08&1.4&0:54:08.53&-72:32:08.7&$\dots$&$\dots$&smc719.10.34369\\ 
92&107&00:54:09.3&-72:41:43&1.4&0:54:09.54&-72:41:43.1&$\dots$&$\dots$&smc105.6.7 (III)\\ 
93&111&00:55:07.7&-72:22:40&0.9&0:55:07.80&-72:22:41.1&$\dots$&$\dots$&smc719.19.88\\ 
94&114&00:56:05.5&-72:00:11&2.0&0:56:05.74&-72:00:11.8&$\dots$&$\dots$&smc719.26.20124\\
95&115&00:56:13.9&-72:30:00&1.0&0:56:13.83&-72:30:01.0&$\dots$&$\dots$&smc719.18.7\\ 
96&116&00:56:14.6&-72:37:56&0.8&0:56:14.67&-72:37:55.4&$\dots$&$\dots$&smc719.09.29033\\ 
97&119&00:57:23.7&-72:23:56&0.8&0:57:23.96&-72:23:56.7&$\dots$&$\dots$&smc719.18.183\\ 
98&121&01:00:30.3&-72:20:33&1.0&1:00:30.03&-72:20:33.5&$\dots$&$\dots$&smc725.07.22332\\ 
99&123&01:00:55.8&-72:23:20&1.0&1:00:55.96&-72:23:21.7&$\dots$&$\dots$&smc725.07.324\\ 
100&124&01:01:47.6&-71:55:51&0.9&1:01:47.90&-71:55:51.1&$\dots$&$\dots$&smc113.5.8 (III)\\ 
101&125&01:01:55.8&-72:32:37&0.6&01:01:55.80&-72:32:36.4&$\dots$&36.4&smc726.30.25187\\ 
102&126&01:01:55.9&-72:10:28&0.9&1:01:56.09&-72:10:28.7&$\dots$&$\dots$&smc725.14.75\\ 
103&128&01:03:31.7&-73:01:44&1.0&1:03:31.51&-73:01:43.5&$\dots$&$\dots$&smc726.21.100\\ 
104&129&01:03:33.6&-72:04:17&1.6&1:03:33.52&-72:04:15.1&$\dots$&$\dots$&smc725.14.4210\\ 
105&133&01:04:35.5&-72:21:47&0.8&1:04:35.44&-72:21:48.6&$\dots$&$\dots$&smc725.05.18123\\ 
106&136&01:06:33.0&-73:15:43&0.6&1:06:32.97&-73:15:42.8&$\dots$&$\dots$&smc111.3.14 (III)\\ 
107&137&01:07:44.5&-72:27:42&0.6&01:07:44.2&-72:27:41.1&$\dots$&$\dots$&smc110.4.19754 (III)\\ 
108&138&01:08:20.2&-72:13:47&0.7&1:08:20.41&-72:13:49.3&$\dots$&$\dots$&smc725.03.14475\\ 
109&139&01:15:03.5&-73:28:19&1.0&1:15:03.41&-73:28:21.4&$\dots$&$\dots$&smc733.22.3637\\ 
110&141&01:19:38.9&-73:30:11&0.7&1:19:39.00&-73:30:11.6&$\dots$&$\dots$&smc733.20.2992\\ 
111&143&01:23:27.5&-73:21:23&1.1&01:23:27.42&-73:21:22.3&$\dots$&$\dots$&smc733.26.24\\
\end{longtable}
\begin{tablenotes}
\item[\textit{Notes}] Source numbers, which we use throughout this paper, increase with increasing spin period and then with increasing RA for sources without measured spin periods. The corresponding source number used by \citetalias{Haberl2016} is indicated, with Source \#2 having two designations thanks to more recent studies finding separately-cataloged sources to be the same system. OGLE IDs are for OGLE IV unless otherwise noted.
We indicate sources with updated X-ray coordinates using the following superscripts: \textsuperscript{a} \citet{2015ATel.8091....1K} \textsuperscript{b} \citet{2014ATel.5772....1K} \textsuperscript{c} \citet{2018ATel12219....1G}.
\end{tablenotes}
\end{tiny}
}

Examining the long-term optical light curves of these sources, we find a range of phenomena. We divide our sample into five types associated with qualitative aspects of their observed light curves and distinguished by quantifiable variability metrics.

\begin{figure}
    \centering
    \includegraphics[width=\columnwidth]{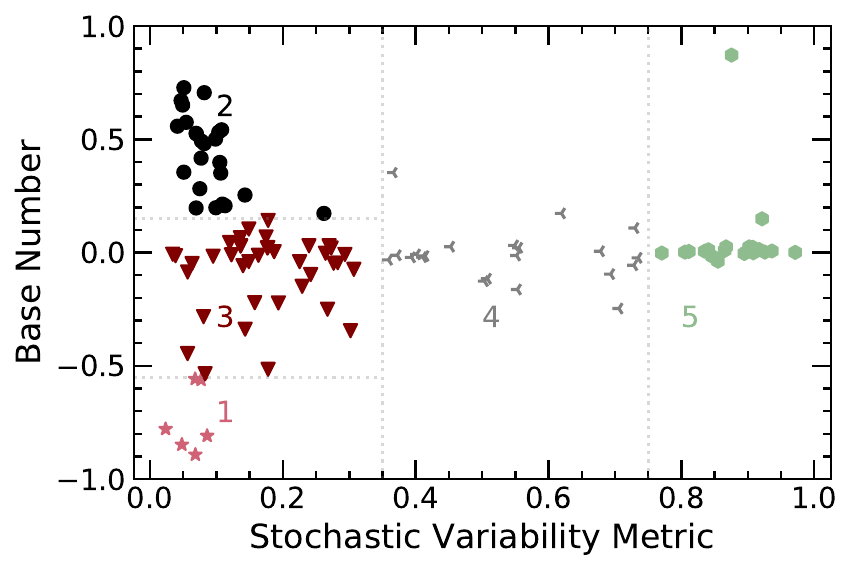}
    \caption{Base number (i.e., $\mathrm{(Max\,I - Median\,I) - (Median\,I - Min\,I)}$) versus the ratio between the standard deviation of the detrended and original I mag light curves for all sources. The gray  dotted lines define the separations between the super-orbital types. The color and shape coding is consistent in all plots. Types 1-5 are represented by pink stars, black circles, maroon triangles, gray crosses, and green hexagons, respectively.
    }
    \label{fig:lbase_stdev}
\end{figure}

Our quantitative distinctions rely on two parameters. The first metric, hereafter referred to as ``base number," is $\mathrm{(Max\,I - Median\,I) - (Median\,I - Min\,I)}$, where I is the I-band magnitude (and thus lower numbers are brighter). 
This metric measures the separation between the median brightness and the extrema. Systems in which there is a ``base'' near minimum brightness with occasional outbursts will have a relatively large negative value of the base number. Systems which generally stay at their bright ends but that have occasional dips will have a large positive value of the base number. Base numbers near zero represent the lack of a consistent level to which the light curve returns, or, in select cases, the potential presence of both a bright and a faint base. 
To protect this quantity from unphysical minima and maxima, we use the third brightest and third faintest points as the ``Min" and ``Max." 
This metric is comparable to the skew of the light curve but better quantifies the presence of an extreme value to which the light curve returns. 

The second metric is a measure of the strength of short time scale stochastic variability relative to coherent variability on super-orbital time scales. We refer to this as the stochastic variability metric, which we define as the ratio of the detrended standard deviation to the standard deviation of the original data. 
Lower values are associated with sources for which the long-term super-orbital modulation is large compared to the stochastic variability. To once again control against unphysical outliers in the original and detrended data, we performed ten and five sigma-clipping, respectively. 
This metric is sensitive to the detrending approach (see Sect. \ref{sec:period}), so we use consistent, reasonable parameters across the sample.

\begin{figure*}
    \centering
    \includegraphics[width=2.1\columnwidth]{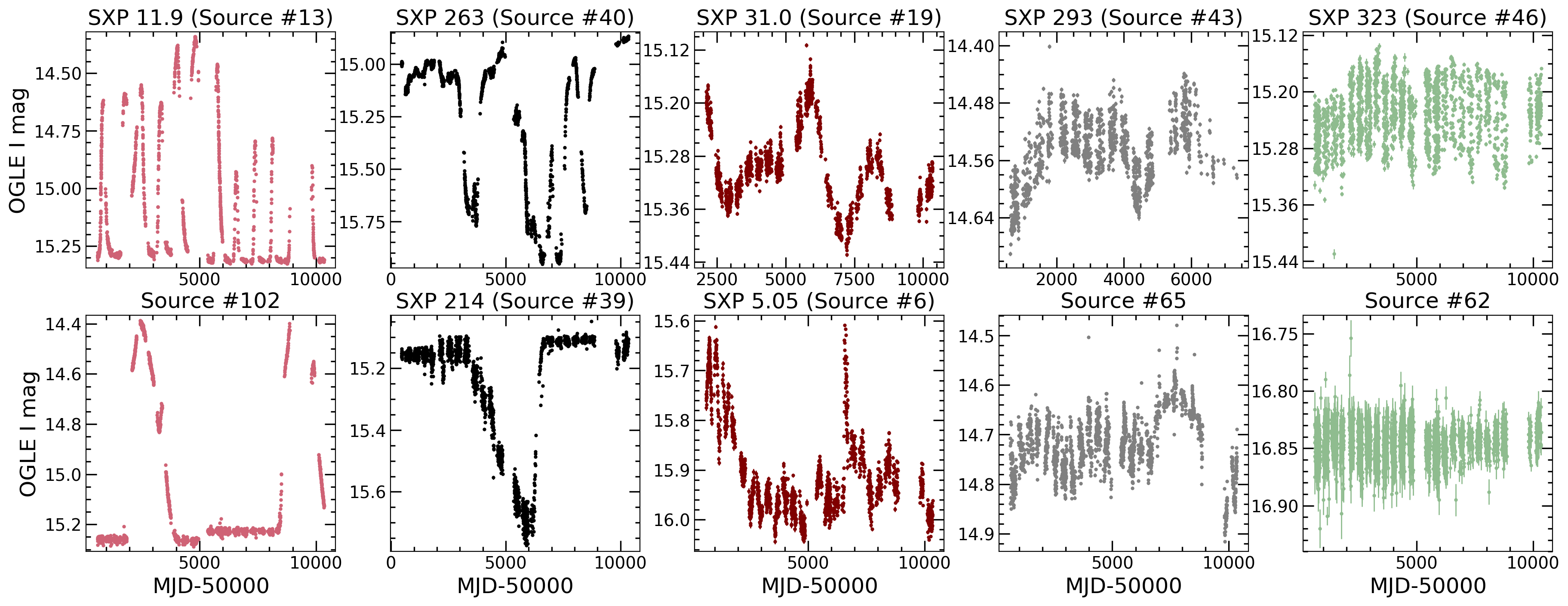}
    \caption{Two example light curves for each of the five source types. The type number increases with each column from left to right. 
}
    \label{fig:rep_types}
\end{figure*}

We plot our sources in the base number versus stochastic variability plane in Fig. \ref{fig:lbase_stdev}. The five types of sources are delineated in that plot and formally defined in Table \ref{t.type_definitions}. 
Although it is not straightforward where to place the cutoff between types, we find it helpful to define discrete types, as long as we keep in mind that the specific location of sources in the parameter space of Fig. \ref{fig:lbase_stdev} is sometimes more illustrative than the discrete type number. 
In Fig. \ref{fig:rep_types}, we show two examples of each of the five light curve morphologies. We use this taxonomy throughout the paper, finding that the types have commonalities beyond those in the definition and the corresponding qualitative super-orbital variability patterns. 

In Figs. \ref{fig:type1gallery}--\ref{fig:type6gallery2}, we show all the light curves of each type, with the light curves within each type ordered by source number.
Here, we provide an overview of each type, and we discuss the physical implications in Sect. \ref{sec:discussion}.

Type 1 sources occupy the low base number and low stochastic variability regime and thus have a coherent minimum brightness with occasional flares. Type 1 sources are the most cohesive type because of their self-similarity in both super-orbital and color-magnitude properties.
In particular, the total I-band variability ranges between 0.7 and 1.1 magnitudes across the sources. The median V$-$I and I-band magnitudes all fall within 0.1 and 1.0 magnitudes of each other, respectively.
We find that none of these sources have strictly periodic super-orbital behavior (Sect. \ref{sec:period}). However, the flares can appear fairly regular, with spacings of $\sim$1000 days for most sources but $\gtrsim$6000 days between the two flares of Source \hyperref[fig:type1gallery]{\#102}. 

Type 2 sources have a well-defined maximum brightness where the sources spend much of their time, interrupted by occasional dips. The Type 2 sources do not simply resemble flipped Type 1 sources as they lack the Type 1 level of self-similarity. 
In particular, some maxima have low variability resembling the Type 1 faint bases, whereas in others the maximum value varies slowly over time.  Furthermore, some Type 2 sources have frequent dips within the OGLE coverage---sometimes with monotonic changes to the dip amplitudes---and others show only one substantial dip. Compared to the Type 1 sources, there is also more spread in their I-band variability ranges (0.35 to 1.16 mag) and median color-magnitude values ($-$0.13 to 0.45 and 13.8 to 15.4 mag). 

Type 3 sources are intermediate between the Type 1 and Type 2 sources, in that their variability is dominated by coherent super-orbital variability, but there is no well-defined maximum or minimum value to which the sources continually return. 

Type 4 and 5 sources are not dominated by super-orbital variability; the dividing line between these categories of sources is somewhat arbitrary, but there do seem to be significant differences between Type 4 and Type 5 sources, as discussed in Sect.\,\ref{sec:discussion}.

Type 1--3 sources have intervals between the flares and dips during which the sources have little coherent variability and thus look similar to Type 4 or Type 5 sources. This raises the possibility that some of our Type 4 and Type 5 sources might someday exhibit more significant coherent variability and thus move into a 
different part of Fig. \ref{fig:lbase_stdev}.
To better understand the potential for such transitions, we analyzed the parts of Type 1--3 light curves with low coherent variability. We divided each light curve of Type 1--3 into 1000-day long pieces and identified segments that fall into the Type 4 or 5 part of Fig. \ref{fig:lbase_stdev}. We subsequently extended and shifted these light curve pieces to identify the full quiescent region.
We find that 84\% of sources in Types 1--3 resemble Type 4 sources for at least 1000 days whereas only 54\% have Type 5 behavior for that long.
Thus it seems likely that some of our Type 4 sources could at some point become as variable as Type 1--3 sources, while Type 5 sources are less likely to leave their quiescence. We further discuss these types of transitions in Sects. \ref{sec:color} and \ref{sec:discussion}.

Past taxonomies of long-term optical Be or BeXRB light curves have been qualitative but with notable overlap with our own classification. To varying extents, \citet{Mennickent2002}, \citet{Schmidtke2013}, and \citet{Kourniotis2014} all focus on the presence of high and low states, seemingly random long-term variability, and (quasi-)periodic outbursts. For instance, our Type 1 sources are similar to hump-like type-1 sources in \citet{Mennickent2002}, the \citet{Kourniotis2014} bumper Be stars, and a subset of the swooper BeXRBs in \citet{Schmidtke2013}. 

\section{Color-magnitude behavior}
\label{sec:color}
\begin{figure}
    \centering
    \includegraphics[width=\columnwidth]{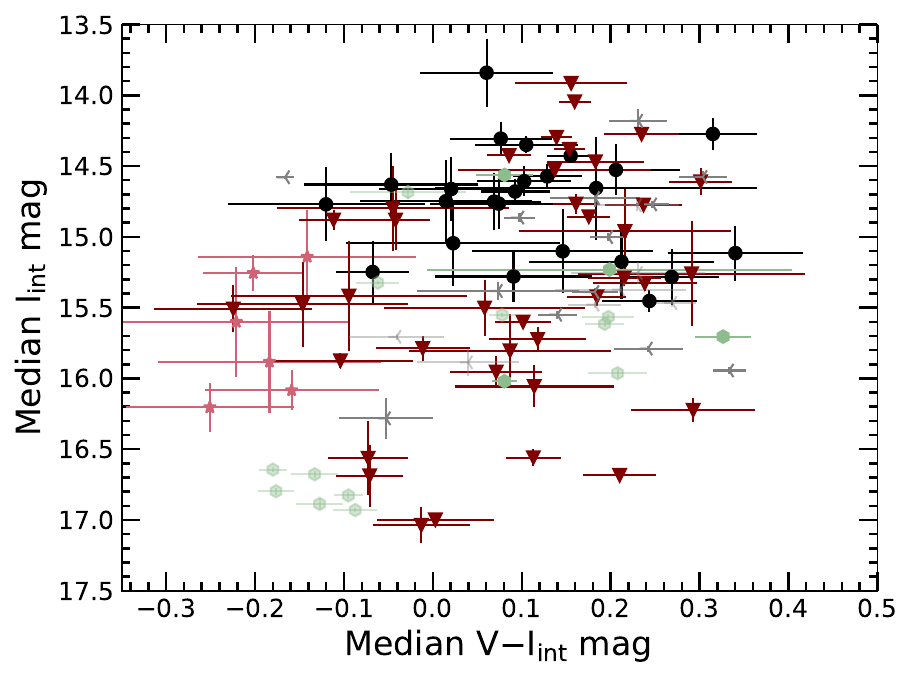}
    \caption{Color-magnitude diagram for all sources with types distinguished as they are in Fig. \ref{fig:lbase_stdev}. Median light curve I and V$-$I magnitudes are used, with error bars representing the standard deviation of each. Translucent points indicate that the corresponding source required manual calibration of OGLE epochs, entailing an additional source of uncertainty in both axes.
    We exclude sources 1, 48, 73 and 92, due to their limited overlap in I- and V-band coverage.
For eight sources (Sources \hyperref[fig:type5gallery1]{63}, \hyperref[fig:type5gallery2]{74}, \hyperref[fig:type5gallery2]{77}, \hyperref[fig:type3gallery3]{79}, \hyperref[fig:type6gallery2]{86}, \hyperref[fig:type3gallery3]{88}, \hyperref[fig:type3gallery3]{91}, and \hyperref[fig:type3gallery3]{98}), there appear to be unphysical measurements during OGLE-II, so these measurements are excluded from the distributions.}
    \label{fig:allcm}
\end{figure}

To better understand the physical processes underlying the observed optical variability, we also analyzed the color-magnitude behavior of the systems.
To create an I versus V$-$I color-magnitude diagram (CMD) for each source, we used linear interpolation for an estimate of I mag values at the times of V-band observations. We chose to interpolate in I-band since the data is more densely sampled.  We show the resulting CMD in Fig. \ref{fig:allcm}, where we plot the median location of each source, along with the standard deviation in both magnitude and color.  

Comparing the positions in the CMD of the different source categories, we find that Type 1 sources are generally fainter and bluer than the Type 2 systems. The Type 3 and Type 4 sources have colors comparable to the Type 2 sources, but Type 3 systems exist at a broad range of magnitudes. In particular, several sources are fainter than the Type 1 objects (Sources \hyperref[fig:type3gallery1]{3}, \hyperref[fig:type3gallery2]{47}, \hyperref[fig:type3gallery2]{64}, \hyperref[fig:type3gallery3]{78}, and \hyperref[fig:type3gallery3]{91}).
At the faint blue end, there is also a noteworthy cluster of Type 5 sources (Sources \hyperref[fig:type6gallery1]{65}, \hyperref[fig:type6gallery1]{67}, \hyperref[fig:type6gallery1]{71}, \hyperref[fig:type6gallery2]{87}, \hyperref[fig:type6gallery2]{94}, and \hyperref[fig:type6gallery2]{109}).

\begin{figure}
    \centering
    \includegraphics[width=\columnwidth]{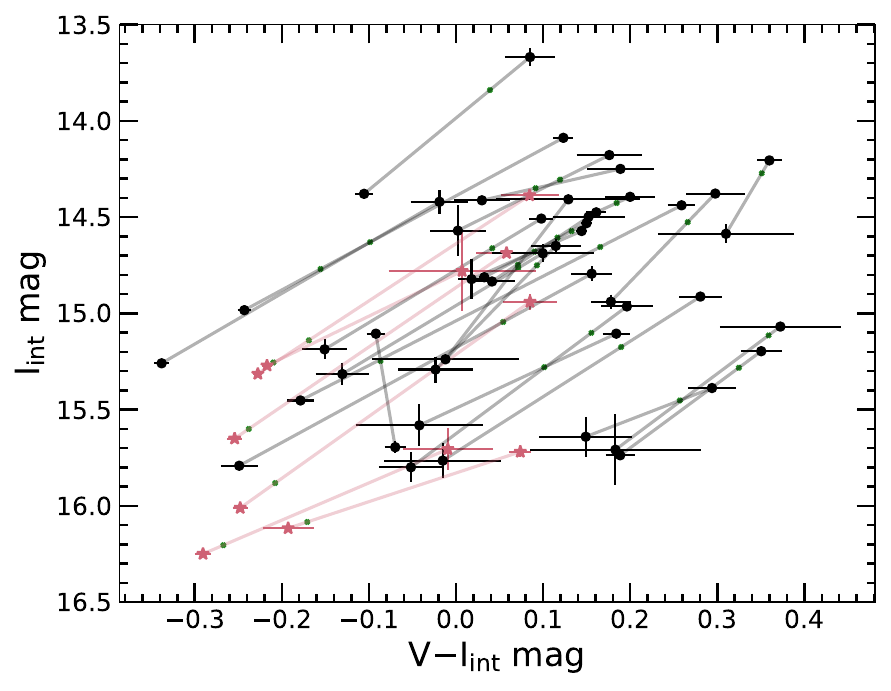}
    \caption{Color-magnitude diagram for Type 1 and 2 sources. The lines connect the median behavior for a given source at its bright and faint ends. Green X's indicate the overall median I magnitude for each source. 
    Because we place these X's along the connecting lines, they do not necessarily mark the median V$-$I.}
    \label{fig:color_change}
\end{figure}

These observations apply only to the median magnitude and color of the systems. To explore the ranges occupied by the sample, we determined the median color and magnitude for the parts of the light curve brighter than the 10th percentile in magnitude and fainter than the 90th percentile. In Fig. \ref{fig:color_change}, we connect these extrema on a CMD for each source in Types 1 and 2. 
We find that the bright end of the I mag range for the Type 1 sources overlaps with the range of magnitudes of the Type 2 sources.  However, we find that a majority of Type 2 sources are redder at their maxima than the Type 1 are at theirs. In other words, the red color of the Type 2 sources is a stronger differentiator than their brightness. Still, one Type 2 source (Source \hyperref[fig:type2gallery1]{\#11}) becomes both as blue and as faint as the Type 1 sources and a few others are nearby (Sources \hyperref[fig:type2gallery1]{2}, \hyperref[fig:type2gallery1]{10}, \hyperref[fig:type2gallery1]{12}, and \hyperref[fig:type2gallery2]{89}). Notably, these Type 2 sources also have potential Type 1-like faint bases in their light curves.

Furthermore, most of the lines in Fig. \ref{fig:color_change} connecting the faint and bright color-magnitude behavior illustrate the common correlation between brightening and reddening. Especially in the middle region of brightness, Type 1 and Type 2 sources show similar redder-when-brighter slopes.
This relationship corresponds to the higher variability amplitudes in I-band than V.

\begin{figure*}
    \centering
    \includegraphics[width=2\columnwidth]{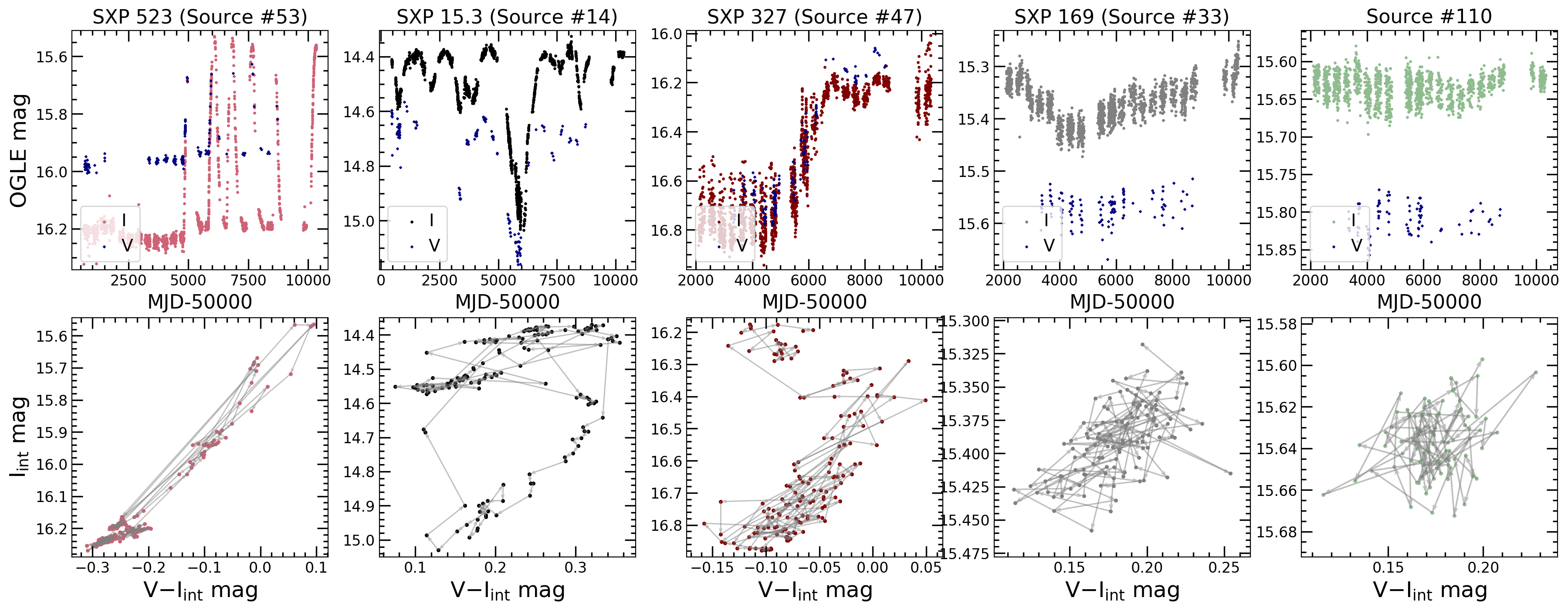}
    \caption{Representative color-magnitude behavior for one source of each super-orbital type, with the type number increasing from left to right. I- and V-band light curves are each plotted with the corresponding (interpolated) I versus V$-$I plot below. Arrows trace the time evolution of the CMD.}
    \label{fig:cm_gal}
\end{figure*}

The color-magnitude behavior also helps our understanding of the potential transitions between types. In Fig. \ref{fig:allcm} we see that the Type 4 systems
occupy parts of the CMD associated with Type 2 systems, and are generally brighter and redder than Type 1 systems. Besides the faint blue group of Type 5 systems, which may not be BeXRBs at all as discussed in Sect. \ref{sec:discussion}, most Type 5s also overlap with Type 2.
Furthermore, the Type 1 source SXP 525 has $\sim$4000 days of approximate quiescence during which it stays in the Type 1 region of the CMD.
Thus, there is overlap in the CMDs between Type 2 bright ends and Type 4 and 5 light curves, whereas the Type 1 sources appear to be more distinct.  This suggests that transitions away from Types 4 and 5 are more likely to result in Type 2 or 3 systems than in Type 1.

Nonetheless, the color-magnitude behavior of Type 1 source SXP \#164 suggests that this system has become Type 2, 3, or 4 in the final few thousand days, although it most recently returned to its faint base level (Fig. \ref{fig:type1gallery}). In particular, rather than descending from one of its few semi-regular brightening events, SXP \#164 continued to brighten and has remained in a brighter and redder state than previously observed.

Representative CMDs for sources of each super-orbital type are shown in Fig. \ref{fig:cm_gal}. The light curves of many sources have a straightforward linear correlation between redness and brightness, with varying levels of scatter by source. Several sources have turnovers or flatten (i.e., change in color while maintaining approximately the same brightness) in either direction at bright values. A few diagrams have significant scatter without a clear relationship, while others maintain the redder-when-brighter relation on the whole but have complex features. 

\begin{figure}
    \centering
    \includegraphics[width=\columnwidth]{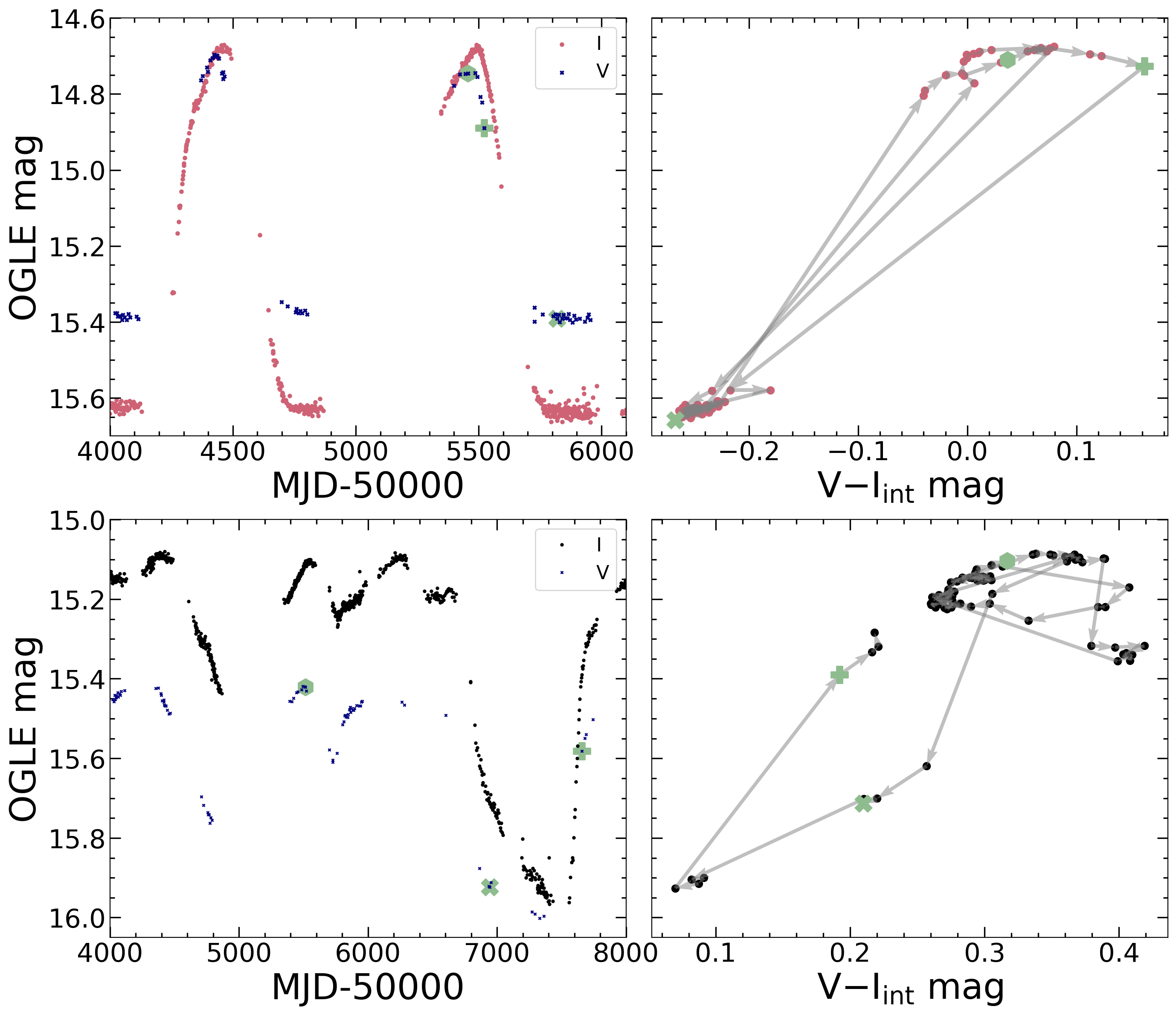}
    \caption{Examples of looping behavior for Type 1 and 2 sources. \textit{Top:} Two peaks of a representative Type 1 light curve (Source \hyperref[fig:type1gallery]{\#103}), with the corresponding color-magnitude behavior shown on the right. 
    The larger green points highlight the correspondence between the two plots at three key stages of the light curve evolution.
    \textit{Bottom:} Partial Type 2 light curve of Source \hyperref[fig:type2gallery2]{\#75} with the corresponding part of the source's CMD on the right.
    }
    \label{fig:loop_examples}
\end{figure}

Loop structures are one such feature. We find evidence for loops in the CMDs of several of our sources and show two examples in Fig. \ref{fig:loop_examples}.
\cite{deWit2006} identified similar loops in single SMC Be stars and modeled this behavior as a transition from an optically thick to optically thin state in the Be disk. 
Similarly, \citet{Haubois2012} modeled Be disk build-up and dissipation, finding that this evolution yields loops in CMDs, with the specific morphology sensitive to inclination.
\citet{Monageng2020} applied this understanding to the loop in the CMD of Source \hyperref[fig:type1gallery]{\#103}.
The \citet{deWit2006} explanation of the looping behavior may also account for another common feature: a decreased slope at bright values, such that systems get redder while staying around the same brightness. In particular, this flattening may arise at the beginning of inside-out clearing, when this change disproportionately affects the bluer emission, leaving the system's overall optical emission redder. We further discuss the loop behavior in Sect. \ref{sec:discussion}.

\section{Periodicities}
\label{sec:period}
By identifying periodicities in optical data, we can connect the various time scales of variability to the system's physical attributes. When orbital periodicities correlate with other properties of the system, we can constrain how the compact object affects the high-mass donor star and the relevance of the orbital architecture to the X-ray behavior.
Shorter periods ($\lesssim$2 days) are usually ascribed to non-radial pulsations of the Be star \citep[e.g.,][]{Rivinius2003}. These pulsations are a promising potential driver for the formation of the Be decretion disk and their beat periods may even explain much longer-term variations \citep[e.g.,][]{Osaki1986,Rivinius2001NRP}. For most HMXB systems, the dominant, highest-amplitude optical variations occur on time scales of thousands of days, with some studies suggesting these super-orbital variations can themselves recur periodically \citep[e.g.,][]{Rajoelimanana2011}.

\begin{figure*}
    \centering
    \includegraphics[width=2\columnwidth]{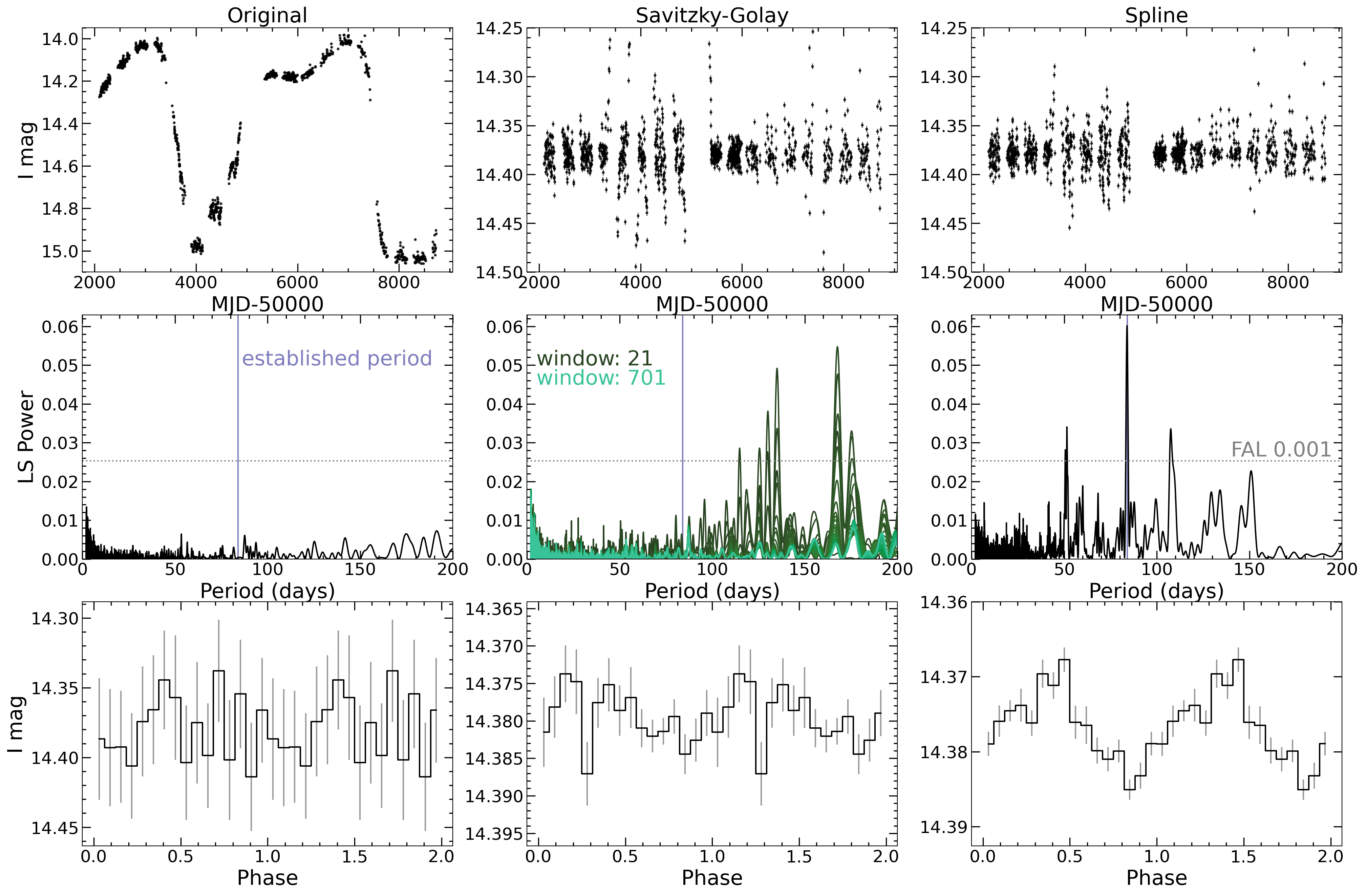}
    \caption{Demonstration of the utility of spline detrending for periodicity searches. The top row includes I magnitude light curves for Source \#2 (SXP 2.16), with the middle column using Savitzy-Golay detrending, and the right using spline detrending with \texttt{wotan} \citep[][]{Hippke2019}. The corresponding periodograms and phase-folded data are shown below these light curves, with the established period from \citetalias{Haberl2016} used in each. 
    We show the Savitzky-Golay detrending using a window of 21 for the light curve and phase-folded data. As noted in Sect. \ref{sec:xrayperiods}, this value is also consistent with X-ray observations of the orbital period. The gray dotted line marks the false alarm level of 0.001.
    }
    \label{fig:spline_ex}
\end{figure*}

Because of the significant super-orbital variability in most of the light curves, Lomb-Scargle periodograms \citep{Lomb1976,Scargle1982} often only identify potential orbital periodicities when the longer-term trends are subtracted from the light curve. We found detrending using a spline fit to be a particularly effective method. Specifically, we use \texttt {rspline} from \texttt{wotan} \citep[][]{Hippke2019} with a window of 200 and a break tolerance of 50. In Fig. \ref{fig:spline_ex} we provide an example comparing the periodicity search using the original data, the spline-detrended data, as well as the light curve detrended using a Savitzky-Golay filter \citep{Savitzky1964}.

Although we focus on a search for potential orbital periodicities, we also search for other periodicities present in the light curves. 
We did separate periodogram searches from 0.1$-$2 days, 2$-$200 days, and 200 days to a third of the light curve duration. 
In addition to this uniform search, we identified the periodogram peak within 10\% of each established period. These aggregated periods largely come from \citetalias{Haberl2016} and \citet{Coe2015}.

In Sect. \ref{sec:xrayperiods}, we leverage systems with orbital periods established using X-ray data to contextualize and inform our approach to our periodicity search with the OGLE data. As we explain in Sect. \ref{sec:confidence}, we perform a search for orbital periods using Lomb-Scargle periodograms and present measurements guiding our confidence in these values. In subsequent subsections we highlight notable findings from the other time scales.
Table 3 includes established, updated, and new periodicities and the corresponding confidence descriptors.

\longtab{
\setlength{\tabcolsep}{3pt} 
\begin{tiny}
\begin{landscape}
\begin{longtable}
{lllllllllllll}
\caption{\label{t.optical}
Optical properties of the sample.}\\
\textbf{Source \#} & \textbf{Type} & \textbf{Base \#} & \textbf{Var. Metric} &  \textbf{Est. Pds} & \textbf{Best Near Est.} & \textbf{Best Pd} & \textbf{10\%ile I} & \textbf{Med. I} & \textbf{90\%ile I} & \textbf{10\%ile V$-$I} & \textbf{Med. V$-$I} & \textbf{90\%ile V$-$I}\\
\hline
\textbf{} & \textbf{} & \textbf{mag} & \textbf{} &  \textbf{days} & \textbf{days} & \textbf{days} & \textbf{mag} & \textbf{mag} & \textbf{mag} & \textbf{mag} & \textbf{mag} & \textbf{mag}\\
\hline
\hline\noalign{\smallskip} 
\endfirsthead
\caption{continued.}\\
\textbf{Source \#} & \textbf{Type} & \textbf{Base \#} & \textbf{Var. Metric} &  \textbf{Est. Pds} & \textbf{Best Near Est.} & \textbf{Best Pd} & \textbf{10\%ile I} & \textbf{Med. I} & \textbf{90\%ile I} & \textbf{10\%ile V$-$I} & \textbf{Med. V$-$I} & \textbf{90\%ile V$-$I}\\
\hline
\textbf{} & \textbf{} & \textbf{mag} & \textbf{} &  \textbf{days} & \textbf{days} & \textbf{days} & \textbf{mag} & \textbf{mag} & \textbf{mag} & \textbf{mag} & \textbf{mag} & \textbf{mag}\\
\hline
\hline\noalign{\smallskip} 
\endhead
\hline
\endfoot
1&5&0.00&0.97&$3.89^{X(a)},3.89^{H}$&$3.89^{X},3.89^{H}$&\textbf{2.05}&13.32&13.38&13.43&$\dots$&$\dots$&$\dots$\\
2&2&0.67&0.05&$82.0^{X(b)}$,$84.0^{H}$&$83.94^{X}$,$83.94^{H}$&$\textbf{83.94}$&14.04&14.18&14.98&-0.24&-0.12&0.06\\
3&2&0.71&0.08&$18.38^{X(c)}$,$18.38^{H}$,$18.6^{C}$&$17.4^{H}$,$17.4^{C}$&127.11&14.67&14.77&15.30&0.00&0.07&0.12\\
4&3&0.07&0.17&$82.37^{H}$,$81.81^{C}$&$81.89^{H}$,$81.89^{C}$&$\textbf{81.89}$&13.88&13.91&14.00&0.08&0.16&0.23\\
5&4&-0.03&0.36&$\dots$&$\dots$&2.23&15.50&15.55&15.61&0.11&0.14&0.16\\
6&3&-0.22&0.19&$17.13^{X(d)}$,$17.13^{H}$,$17.2^{C}$&$17.34^{H}$,$17.34^{C}$&5.72&15.74&15.94&16.0&-0.01&0.07&0.13\\
7&2&0.50&0.10&$21.9^{X(c)}$,$21.9^{H}$,$22.0^{C}$&$23.97^{H}$,$23.97^{C}$&108.61&14.51&14.66&15.07&-0.13&0.02&0.10\\
8&3&0.07&0.14&$44.92^{X}$,$44.92^{H}$,$44.8^{C}$&$44.91^{H}$,$44.91^{C}$&$\textbf{44.91}$&14.72&14.77&14.83&0.13&0.16&0.20\\
9&3&-0.15&0.23&$40.03^{H}$,$35.61^{C}$&$39.83^{H}$,$39.05^{C}$&$\textbf{39.83}$&15.77&15.88&15.92&-0.17&-0.10&0.04\\
10&2&0.36&0.05&$28.47^{X}$,$28.51^{H}$&$28.48^{X}$,$28.48^{H}$&112.33&14.44&14.66&15.24&-0.11&0.01&0.14\\
11&2&0.42&0.08&$36.3^{X(e)}$,$36.3^{H}$,$36.3^{C}$&$36.18^{H}$,$36.18^{C}$&103.80&14.85&15.08&15.57&-0.23&0.02&0.17\\
12&2&0.21&0.11&$35.6^{H}$,$35.4^{C}$&$35.63^{H}$,$35.63^{C}$&35.63&14.47&14.72&15.28&-0.20&-0.05&-0.01\\
13&1&-0.81&0.09&$\dots$&$\dots$&146.08&14.57&15.19&15.31&-0.23&-0.14&0.08\\
14&2&0.48&0.08&$74.51^{H}$,$74.3^{C}$&$74.32^{H}$,$74.32^{C}$&$\textbf{74.32}$&14.39&14.48&14.83&0.12&0.21&0.32\\
15&2&0.40&0.11&$36.43^{H}$&$36.43^{H}$&$\textbf{36.43}$&14.72&14.78&15.11&-0.08&0.07&0.10\\
16&3&-0.06&0.14&$17.73^{X}$,$17.79^{H}$,$17.79^{C}$&$17.96^{X}$,$17.96^{H}$,$17.96^{C}$&130.79&14.95&15.42&15.94&0.05&0.29&0.40\\
17&3&0.14&0.18&$75.97^{H}$&$83.40^{H}$&83.40&14.01&14.04&14.12&0.13&0.16&0.18\\
18&3&0.03&0.27&$22.5^{H}$,$22.5^{C}$&$22.51^{H}$,$22.51^{C}$&11.26&15.6&15.70&15.82&0.06&0.12&0.20\\
19&3&-0.04&0.23&$90.5^{H}$,$90.4^{C}$&$90.42^{H}$,$90.42^{C}$&$\textbf{90.42}$&15.19&15.30&15.35&0.19&0.22&0.29\\
20&2&0.25&0.14&$137.36^{X}$,$137.36^{H}$,$137.4^{C}$&$137.76^{H}$,$137.76^{C}$&$\textbf{137.76}$&14.59&14.66&14.81&0.03&0.09&0.14\\
21&2&0.28&0.07&$122.1^{X}$,$62.1^{H}$,$122.0^{C}$&$114.88^{X}$,$60.91^{H}$,$114.90^{C}$&114.90&15.11&15.32&15.56&-0.03&0.09&0.17\\
22&4&0.11&0.73&$110.6^{H}$,$111.0^{C}$&$110.88^{H}$,$110.88^{C}$&$\textbf{55.33}$&15.69 (c)&15.75&15.82&0.00&0.04&0.06\\
23&3&-0.10&0.24&$61.6^{X}$,$33.38^{H}$,$33.3^{C}$&$61.09^{X}$,$33.41^{H}$,$33.41^{C}$&$\textbf{33.41}$&16.64&16.68&16.77&0.18&0.21&0.25\\
24&3&-0.01&0.04&$362.3^{X}$,$362.3^{H}$,$362.0^{C}$&$362.51^{H}$,$362.51^{C}$&103.33&14.74&15.07&15.5&0.03&0.22&0.34\\
25&4&-0.02&0.73&$87.6^{X}$,$88.3^{H}$,$88.0^{C}$&$88.31^{X}$,$88.32^{H}$,$88.32^{C}$&$\textbf{88.32}$&14.73&14.77&14.80&0.22&0.25&0.27\\
26&3&-0.05&0.28&$21.95^{H}$,$21.9^{C}$&$21.96^{H}$,$21.96^{C}$&$\textbf{21.96}$&15.5&15.60&15.63&0.07&0.10&0.16\\
27&3&-0.02&0.09&$103.6^{X}$,$103.6^{H}$,$125.0^{C}$&$101.06^{H}$,$122.49^{C}$&122.49&15.87&16.06&16.26&-0.02&0.11&0.24\\
28&3&-0.09&0.06&$197.0^{H}$,$197.0^{C}$&$185.88^{H}$,$185.88^{C}$&$\dots$&15.36&15.73&16.12&-0.10&0.09&0.19\\
29&3&-0.01&0.03&$426.0^{H}$&$402.36^{H}$&$\dots$&14.87&15.14&15.52&-0.24&-0.22&-0.21\\
30&2&0.52&0.07&$\dots$&$\dots$&72.49&15.39&15.44&15.59&0.19&0.24&0.31\\
31&3&-0.34&0.14&$100.3^{C}$&$101.07^{C}$&$\textbf{51.39}$&16.01&16.55&16.74&-0.14&-0.07&-0.03\\
32&1&-0.89&0.07&$29.75^{H}$&$28.37^{H}$&121.29&15.31&15.95&16.03&-0.24&-0.18&0.08\\
33&4&-0.02&0.41&$68.54^{X}$,$68.54^{H}$,$68.6^{C}$&$68.64^{H}$,$68.64^{C}$&68.64&15.33&15.37&15.42&0.15&0.18&0.21\\
34&2&0.20&0.10&$70.42^{X(f)}$,$67.88^{H}$,$68.8^{C}$&$68.59^{X}$,$68.59^{H}$,$68.59^{C}$&22.86&14.26&14.35&14.46&0.04&0.11&0.18\\
35&4&-0.01&0.37&$87.2^{C}$&$87.34^{C}$&87.34&14.79 (c)&14.80&14.82&0.20&0.23&0.23\\
36&4&0.35&0.37&$60.623^{G}$&$60.69^{G}$&\textbf{60.69}&16.17&16.29&16.48&-0.11&-0.01&0.04\\
37&2&0.21&0.11&$71.98^{H}$&$72.30^{H}$&36.28&14.53&14.58&14.75&0.04&0.13&0.15\\
38&3&-0.01&0.16&$224.0^{H}$,$224.6^{C}$&$226.95^{H}$,$226.95^{C}$&$\textbf{113.64}$&15.26&15.33&15.42&0.21&0.24&0.26\\
39&2&0.53&0.10&$29.91^{H}$&$29.87^{H}$&14.93&15.11&15.16&15.66&-0.10&-0.07&-0.01\\
40&5&0.00&0.89&$49.06^{H}$,$49.2^{C}$&$51.73^{H}$,$51.73^{C}$&$\dots$&16.01&16.02&16.03&0.07&0.08&0.09\\
41&3&0.00&0.19&$64.8^{X}$,$64.8^{H}$,$127.3^{C}$&$63.58^{H}$,$126.87^{C}$&$\textbf{126.87}$&15.29&15.39&15.47&0.14&0.18&0.22\\
42&4&0.02&0.56&$151.0^{X}$,$59.77^{H}$&$141.47^{X}$,$59.73^{H}$&$\textbf{59.73}$&14.49&14.54&14.62&0.26&0.30&0.33\\
43&2&0.65&0.05&$\dots$&$\dots$&127.48&15.00&15.11&15.76&-0.02&0.15&0.25\\
44&5&0.00&0.77&$344.0^{H}$,$520.0^{C}$&$338.54^{H}$,$503.70^{C}$&$\dots$&15.56 (c)&15.57&15.58&0.17&0.20&0.20\\
45&1&-0.56&0.07&$1163.0^{H}$&$1163.94^{H}$&21.70&15.81&16.09&16.11&-0.18&-0.16&0.05\\
46&5&0.03&0.87&$116.6^{X}$,$116.6^{H}$&$104.95^{H}$&2.45&15.18&15.23&15.29&0.15&0.20&0.24\\
47&3&-0.25&0.27&$45.9^{H}$,$45.99^{C}$&$45.94^{H}$,$45.94^{C}$&$\textbf{45.94}$&16.21&16.65&16.83&-0.11&-0.07&-0.02\\
48&3&-0.04&0.28&$63.83^{M}$&$63.80^{M}$&\textbf{63.80}&14.67&14.68&14.69&\dots&\dots&\dots\\
49&3&-0.53&0.08&$\dots$&$\dots$&143.60&15.04&15.6&15.72&-0.29&-0.15&0.01\\
50&3&-0.45&0.06&$94.4^{H}$,$93.0^{C}$&$100.64^{H}$,$100.64^{C}$&100.64&14.71&15.40&15.64&-0.23&-0.09&0.09\\
51&3&-0.05&0.06&$74.96^{H}$,$75.0^{C}$&$74.02^{H}$,$74.02^{C}$&$\textbf{74.02}$&15.32&15.53&15.88&-0.17&0.06&0.17\\
52&3&0.02&0.18&$265.3^{X}$,$265.3^{H}$&$238.72^{H}$&32.80&14.72&14.76&14.81&0.16&0.24&0.28\\
53&1&-0.56&0.08&$\dots$&$\dots$&129.52&15.86&16.20&16.24&-0.28&-0.25&-0.07\\
54&4&0.01&0.68&$151.8^{X}$,$151.0^{H}$,$152.4^{C}$&$153.61^{X}$,$153.62^{H}$,$153.62^{C}$&81.20&15.77&15.80&15.83&0.18&0.24&0.28\\
55&2&0.57&0.05&$\dots$&$\dots$&70.96&14.51&14.61&14.84&0.05&0.10&0.17\\
56&5&0.02&0.91&$412.0^{H}$,$412.0^{C}$&$380.10^{H}$,$380.10^{C}$&2.15&15.67&15.70&15.74&0.29&0.33&0.36\\
57&4&-0.06&0.73&$\dots$&$\dots$&3.26&15.46 (c)&15.48&15.51&0.15&0.18&0.21\\
58&3&-0.34&0.30&$389.9^{X}$,$390.0^{H}$,$394.0^{C}$&$392.61^{X}$,$392.33^{H}$,$392.33^{C}$&$\textbf{393.33}$&14.51&14.62&14.72&0.25&0.30&0.32\\
59&4&-0.01&0.55&$91.5^{H}$&$90.94^{H}$&3.74&15.92&15.94&15.97&0.32&0.33&0.35\\
60&2&0.54&0.11&$101.94^{H}$,$101.4^{C}$&$102.17^{H}$,$102.17^{C}$&$\textbf{102.17}$&14.20&14.33&14.49&0.03&0.08&0.19\\
61&4&-0.16&0.55&$656.0^{C}$&$661.17^{C}$&$\textbf{655.8}$&14.14&14.18&14.21&0.22&0.23&0.25\\
62&5&0.01&0.87&$26.17^{X(g)}$,$26.17^{B}$&$26.17^{X}$,$26.17^{B}$&$\textbf{26.17}$&14.52&14.56&14.60&0.04&0.08&0.12\\
63&4&-0.02&0.41&$\dots$&$\dots$&2.69&15.32 (c)&15.35&15.37&$\mathrm{0.21_{III,IV}}$&0.21&0.24\\
64&3&-0.22&0.16&$\dots$&$\dots$&$\textbf{102.87}$&16.75&17.00&17.07&-0.04&-0.01&0.03\\
65&5&0.01&0.91&$\dots$&$\dots$&$\dots$&16.83 (c)&16.84&16.86&-0.11&-0.10&-0.08\\
66&5&-0.02&0.85&$\dots$&$\dots$&$\dots$&15.54 (c)&15.55&15.56&0.07&0.08&0.09\\
67&5&0.01&0.92&$\dots$&$\dots$&$\dots$&16.67 (c)&16.69&16.70&-0.14&-0.13&-0.11\\
68&4&-0.02&0.39&$414.91345^{H}$&$376.31^{H}$&84.15&14.64&14.72&14.78&0.10&0.18&0.24\\
69&3&0.10&0.15&$\dots$&$\dots$&2.17&14.24&14.27&14.35&0.15&0.23&0.26\\
70&3&-0.01&0.15&$80.1^{H}$,$40.1^{C}$&$80.82^{H}$,$40.37^{C}$&20.10&16.12&16.24&16.33&0.23&0.29&0.41\\
71&5&0.01&0.94&$\dots$&$\dots$&$\dots$&16.91(c)&16.92&16.94&-0.09&-0.07&-0.05\\
72&3&0.03&0.14&$413.0^{H}$&$411.82^{H}$&411.82&14.39&14.43&14.50&0.04&0.09&0.11\\
73&5&0.87&0.88&$\dots$&$\dots$&$\dots$&19.41 (c)&19.62&19.89&$\dots$&$\dots$&$\dots$\\
74&4&-0.12&0.51&$\dots$&$\dots$&$\dots$&15.71 (c)&15.72&15.73&$\mathrm{-0.05_{III,IV}}$&-0.04&-0.02\\
75&2&0.73&0.05&$\dots$&$\dots$&96.37&15.06&15.12&15.57&0.23&0.34&0.40\\
76&5&0.02&0.90&$\dots$&$\dots$&$\textbf{2.17}$&15.93 (c)&15.96&16.00&0.17&0.21&0.23\\
77&4&-0.01&0.40&$\dots$&$\dots$&115.92&15.46 (c)&15.48&15.50&$\mathrm{0.23_{III,IV}}$&0.27&0.27\\
78&3&-0.07&0.31&$\dots$&$\dots$&5.61&16.47&16.56&16.60&0.08&0.11&0.16\\
79&3&0.00&0.26&$\dots$&$\dots$&117.37&14.34&14.38&14.40&$\mathrm{0.12_{III,IV}}$&0.15&0.17\\
80&2&0.49&0.08&$\dots$&$\dots$&21.04&15.20&15.30&15.72&0.20&0.27&0.34\\
81&2&0.53&0.07&$\dots$&$\dots$&104.80&14.94&15.16&15.64&0.06&0.21&0.31\\
82&5&0.00&0.91&$\dots$&$\dots$&$\dots$&15.32 (c)&15.33&15.34&-0.06&-0.06&-0.04\\
83&3&-0.01&0.12&$\dots$&$\dots$&78.29&14.77&14.82&14.87&0.15&0.18&0.22\\
84&4&0.03&0.55&$17.541^{S}$&17.55&$\textbf{17.55}$&14.81&14.86&14.91&0.08&0.10&0.13\\
85&3&-0.28&0.08&$\dots$&$\dots$&59.41&14.25&14.82&15.08&-0.19&-0.04&0.16\\
86&5&-0.04&0.86&$\dots$&$\dots$&$\dots$&17.19&17.21&17.24&$\mathrm{0.89_{III,IV}}$&0.96&1.00\\
87&5&0.00&0.90&$\dots$&$\dots$&$\dots$&16.80 (c)&16.81&16.83&-0.21&-0.18&-0.17\\
88&3&0.03&0.24&$\dots$&$\dots$&$\textbf{35.43}$&14.50&14.53&14.57&$\mathrm{0.12_{III,IV}}$&0.14&0.19\\
89&2&0.56&0.04&$\dots$&$\dots$&127.16&14.49&14.68&15.42&-0.18&0.18&0.27\\
90&5&0.01&0.81&$\dots$&$\dots$&$\dots$&13.30 (c)&13.30&13.31&0.67&0.68&0.69\\
91&3&-0.01&0.29&$\dots$&$\dots$&$\dots$&16.93&16.99&17.05&$\mathrm{-0.02_{III,IV}}$&0.00&0.02\\
92&4&0.17&0.62&$\dots$&$\dots$&78.66&12.43&12.59&12.86&$\dots$&$\dots$&$\dots$\\
93&4&-0.10&0.69&$\dots$&$\dots$&2.05&14.56&14.58&14.59&-0.18&-0.17&-0.15\\
94&5&0.01&0.84&$\dots$&$\dots$&$\dots$&16.94 (c)&16.96&16.97&-0.19&-0.13&-0.15\\
95&2&0.17&0.26&$72.231^{H}$&$72.30^{H}$&$\textbf{72.30}$&14.40&14.44&14.49&0.13&0.16&0.20\\
96&2&0.20&0.07&$\dots$&$\dots$&77.10&14.22&14.32&14.51&0.25&0.31&0.36\\
97&3&-0.52&0.18&$\dots$&$\dots$&$\textbf{74.97}$&14.78&14.88&14.91&-0.13&-0.11&-0.06\\
98&3&0.02&0.27&$\dots$&$\dots$&$\dots$&14.45&14.47&14.50&$\mathrm{0.16_{III,IV}}$&0.18&0.20\\
99&3&0.04&0.12&$\dots$&$\dots$&7.89&15.64&15.77&15.90&-0.07&-0.01&0.07\\
100&3&0.02&0.18&$\dots$&$\dots$&105.07&14.28&14.34&14.40&0.12&0.14&0.16\\
101&2&0.35&0.11&$36.41^{H}$&$35.79^{H}$&35.79&13.68&13.85&14.33&-0.10&0.06&0.11\\
102&1&-0.78&0.02&$\dots$&$\dots$&$\dots$&14.54&15.23&15.27&-0.22&-0.20&-0.15\\
103&1&-0.85&0.05&$\dots$&$\dots$&$\dots$&14.74&15.60&15.64&-0.25&-0.22&0.05\\
104&5&0.15&0.92&$\dots$&$\dots$&$\dots$&19.08&19.18&19.28&-0.25&-0.17&-0.07\\
105&4&0.03&0.45&$\dots$&$\dots$&4.06&14.97&15.00&15.04&0.18&0.20&0.23\\
106&3&-0.04&0.15&$\dots$&$\dots$&120.39&14.75&15.07&15.30&-0.10&-0.04&0.00\\
107&4&-0.25&0.71&$\dots$&$\dots$&$\textbf{1179}$&15.23 (c)&15.27&15.30&0.16&0.20&0.25\\
108&5&0.00&0.81&$\dots$&$\dots$&2.14&14.67 (c)&14.68&14.69&-0.04&-0.03&-0.01\\
109&5&0.00&0.93&$\dots$&$\dots$&$\dots$&16.64 (c)&16.66&16.67&-0.21&-0.18&-0.18\\
110&5&0.01&0.84&$\dots$&$\dots$&5.18&15.61 (c)&15.63&15.66&0.16&0.19&0.20\\
111&4&-0.13&0.50&$119.8^{H}$&$119.62^{H}$&\textbf{119.64}&15.34&15.40&15.47&-0.11&0.07&0.13\\
\end{longtable}
\begin{tablenotes}
\item[\textit{Notes}] Superscripts indicate the origin of the established period, where \textit{X}, \textit{H}, \textit{C}, \textit{B}, \textit{G}, \textit{M}, and \textit{S} designate periods from X-ray data, and aggregated in \citetalias{Haberl2016}, \citet{Coe2015}, \citet{Bird2012}, \citet{Gaudin2024}, \citet{Maitra2023}, and \citet{Sarraj2012}, respectively. 
The X-ray period is from \citet{Galache2008} unless otherwise noted in the superscript, where (a) is \citet{Schreier1972}, (b) is \citet{Vasilopoulos2017}, (c) is \citet{Townsend2011}, (d) is \citet{Coe2014}, (e) is \citet{Townsend2011a}, (f) is \citet{Schurch2009}, and (g) is \citet{Carpano2017}.
We report the periodogram peaks within 10\% of each established period. The ``Best Pd" is the overall peak in the 2$-$200 day periodogram for the detrended light curve, with the exception of eclipsing source SXP 182 and the longer-period sources.   
We do not further investigate whether our ``Best Pd" is preferred over any periods established in the literature.
While we report a best period for any source with a peak above a false alarm level of 0.001, only the bolded periods are ``confident" as defined in Sect. \ref{sec:confidence}. Even then, there is a chance that this strong signal is not orbital in origin, especially if it is an alias of a system periodicity outside of the search range. 
For details on the periodicity search, see Sect. \ref{sec:period}.
I-band values use the original I mag data whereas V$-$I calculations use V mag data and I-band linearly interpolated at the times of V mag observations. In the tenth percentile I magnitude column, we mark with (c) sources with OGLE IV manually calibrated to the median I magnitude of OGLE II and OGLE III. For sources with unphysical OGLE II V-band measurements, we indicate in the $90^{th}$ percentile column that we only use III and IV for color measurements. For the four sources without reliable color measurements, we do not include any color information.
\end{tablenotes}
\end{landscape}
\end{tiny}
}

\subsection{Learning from sources with orbital periods from X-ray data}
\label{sec:xrayperiods}
It is notoriously difficult to identify physically meaningful periodicities in optical data of HMXBs \citep[e.g.,][]{Bird2012}. Complicating the search are uneven sampling and the potential presence of multiple periodic signals (i.e., pulsations of the Be star, orbital period, QPOs, and super-orbital modulations). These (quasi-)periodicities interact to cause beat periods and aliasing. Furthermore, the amplitudes associated with the orbital period can be comparable to other short-term changes. Because of these issues, the corresponding X-ray data is typically more reliable for identifying the orbital period. Therefore, to explore the accuracy of optical period search, we analyze the optical periodicity of sources in our sample that have measured orbital periods from X-ray data.  In this section we highlight the findings that affect how we interpret periodicity searches for the entire sample.

For each of the 24 sources with an orbital period established with X-ray data, we analyzed the periodogram for the detrended optical data, along with the phase-folded light curves for each established X-ray or optical period, the overall periodogram peak, and the strongest peaks within 10\% of each established period. 
We find seven sources (Sources \hyperref[fig:type2gallery1]{2}, \hyperref[fig:type3gallery1]{8}, \hyperref[fig:type2gallery1]{20}, \hyperref[fig:type2gallery1]{21}, \hyperref[fig:type5gallery1]{25}, \hyperref[fig:type5gallery1]{33}, and \hyperref[fig:type6gallery1]{62}) that recover a value within 10\% of the reported X-ray period as the overall periodogram peak, of which five are well within 1\%. 
For eleven additional sources, the highest periodogram peak is likely a harmonic of the reported X-ray period. This is not unexpected, since the periodogram typically prefers a smooth sinusoidal phase-folded light curve, whereas the expected shape of the orbital light curve is a ``fast rise exponential decay" \citep[e.g.,][]{Bird2012}.
In particular, ten of these 24 sources (Sources \hyperref[fig:type6gallery1]{1}, \hyperref[fig:type2gallery1]{3}, \hyperref[fig:type3gallery1]{6}, \hyperref[fig:type2gallery1]{7}, \hyperref[fig:type2gallery1]{10}, \hyperref[fig:type3gallery1]{23}, \hyperref[fig:type2gallery1]{34}, \hyperref[fig:type3gallery2]{41}, \hyperref[fig:type3gallery2]{52}, and \hyperref[fig:type3gallery2]{58}) have best periods within 10\% of a harmonic of the X-ray period, and an eleventh (Source \hyperref[fig:type2gallery1]{\#11}) still could fall into this category given its X-ray period of 36.3 $\pm$ 0.4 days and its LS best of 103.8 days.   
Of these eleven sources, four have a value above a false alarm level (FAL) of 0.001 within 10\% of the X-ray period. 
In only one of these four cases (Source \hyperref[fig:type3gallery2]{\#41}) is the harmonic clearly the preferable period, according to the amplitude of the phase-folded profiles. This peak is within a day of the \citet{Coe2015} reported optical period. 
On the other hand, while SXP 756 has an overall peak at 98.3 days (even with a maximum search period of 500 days), the phase-folded ranges and visual inspection of the light curve both clearly demonstrate that the true period is within the reported uncertainty of the X-ray value (389.9 $\pm$ 7.0 days). 

The remaining six sources (Sources \hyperref[fig:type3gallery1]{16}, \hyperref[fig:type3gallery1]{24}, \hyperref[fig:type3gallery1]{27}, \hyperref[fig:type5gallery1]{42}, \hyperref[fig:type6gallery1]{46}, \hyperref[fig:type5gallery1]{54}) have non-harmonic periodogram peaks that are not within 10\% of the X-ray values. Two of these peaks (Sources 27 and 42) are close to previously reported optical periods. Again, these eight values seem to optimize phase-folded range, smooth phase-folded shape, or both, but only two (59.73 days and 2.45 days for Sources 42 and 46, respectively) look markedly superior to the X-ray periods. Of these two, one is clearly stronger than the X-ray period of 151 days while the other is a short periodicity for a source without phase-folded support for the X-ray period. Thus, while only a minority of sources (7/24) produced maxima of the periodogram that were the same as the X-ray periods, 75\% of the sources (18/24) yielded either the same period or a harmonic. Of the other six, only two generated folded light curves that are clearly stronger than those associated with the measured X-ray period.

Two examples demonstrate the limits of the X-ray approach to finding a HMXB orbital period. First, \citet{Vasilopoulos2017} found that SXP 2.16 has close but inconsistent X-ray and optical best periods. In particular, with Lomb-Scargle periodograms on \swift/BAT and OGLE data of the source, they detected periods of 82.0 $\pm$ 0.3 and 83.6 $\pm$ 0.2 days, respectively. \citet{Boon2017} also noted this discrepancy, identifying an X-ray period of 82.5 $\pm$ 0.7 days, even when using additional archival \swift data. However, by repeating the X-ray analysis but with the more recent \swift data included, we see that the Lomb-Scargle periodogram peaks at $\sim$83.5 days, consistent with the optical finding. This example highlights both that the reported X-ray period errors can be under-estimated and that the optical approach can occasionally point to a better value.

Second, BeXRB EXO 2030+375 has an orbital period of 46 days \citep{Wilson2005} but during its 2021 giant X-ray outburst, the peaks were separated by double the orbital period\footnote{\url{gammaray.msfc.nasa.gov/gbm/science/pulsars/lightcurves/exo2030.html}}. Especially for sources with X-ray orbital periods estimated using only outbursts, periodogram analysis of X-ray light curves may recover harmonics of the true orbital period as the best period.

Thus, the peak of the optical periodogram is often, but not always, indicative of the true orbital period as determined by the more reliable X-ray data. In particular, the LS periodogram can favor harmonics over periods consistent with the X-ray,  sometimes preferring smoother, more sinusoidal profiles over whichever period and profile maximizes variability amplitude. 
With these findings in mind, we perform periodogram searches and confidence evaluations for the remaining sources in Sect. \ref{sec:confidence}. In this paper, we focus on the overall LS periodogram peaks, but future work should evaluate each significant peak or start with a consistency-driven method.

\subsection{Periodogram search and confidence metrics}
\label{sec:confidence}
Informed by our analysis of sources with X-ray orbital periods (Sect. \ref{sec:xrayperiods}), we found the detrended Lomb-Scargle 2$-$200 day peak for each source. Using qualitatively determined cutoffs of three continuous confidence measures, we categorized the periods into three broad confidence levels. For sources in the lowest confidence level, which have no periodogram signal above a false alarm level (FAL) of 0.001, we do not report any potential periodicities. The sources in the highest confidence level have an LS peak that meets the following criteria, which we discuss in more detail below:
\begin{enumerate}
    \item Using 16 phase bins, the phase-folded bins' range (maximum to minimum bin) divided by their mean standard error exceeds nine.
    \item Using a rolling periodogram of 300 points in steps of 50 points, at least a third of the periodogram peaks are within 5\% of the entire light curve's periodogram peak.
    \item Across separate ten-bin phase-folded profiles of $\sim$300 points each, the mean phase change of the phase-folded peak never exceeds two bins.
\end{enumerate}

In these periodogram searches, we used a maximum period of 200 days, although some systems might have longer orbital periods. This choice enabled us to avoid the aliasing effects near one year. However, for the three sources discussed in Sect. \ref{sec:longperiod}, we adopted a longer value because their sharp peaks are visually obvious, even though they present difficulty for the Lomb-Scargle periodogram. 
Similarly, the periodogram peak for SXP 182 is at 30.3 days, despite the clear, sharp eclipses separated by 60.623 days, as the NS accretion disk blocks light from the high-mass donor star \citep{Gaudin2024WD}.
We therefore provide best values for these four systems using manual identification and refining using an optimization of phase-folded range.
Still, the corresponding 2$-$200 day periodograms identified harmonics of the true periods as the peak. 

After first performing the 2$-$200 day search without referencing the established optical periods that are present for 61 of the 111 
sources, we repeated our steps in targeted searches within 10\% of these previously reported periods. We present these period values and confidence metrics in Table \ref{t.optical}. 
We emphasize that we simply report the 2$-$200 day periodogram peak as the ``Best Period" for all but the four sources mentioned above and we do not perform further analysis to compare this period to previously established or targeted ones. Especially in the case of harmonics, it is possible that the true orbital period is not the periodogram peak, but we expect the ``confident" periods to at least be harmonics of the true ones.
Here, we detail the quantities and cutoffs for the three confidence classes before comparing the unbiased and targeted searches. In Sect. \ref{sec:discussion}, we note the connections between the period values and confidences and other quantities explored in this work.

First, we deemed a source entirely ``aperiodic" if it has no detrended 2$-$200 day periodogram peak above an FAL of 0.001. This threshold matched our qualitative interpretation of the unconvincing phase-folded profiles for these sources. In Table \ref{t.optical}, we do not report any periodicity for these 21 sources. 

For sources with the periodogram peak more significant than this FAL cutoff, we made three cuts in order to identify the ``confident" category, which initially included 30 systems, although we made slight alterations described below that yielded a final category of 33 sources. 
We use three discrete cuts, but each cut uses a continuous value; other thresholds for the cutoffs are certainly possible and these quantities may be informative beyond this use.
For systems with false alarm values below 0.001, we report periodicities even for the sources that fail to meet the additional criteria because some may simply lack sufficient support with the current data or approach and some may represent relevant time scales even if they are not orbital.

The first confidence cut focuses on the overall phase-folded strength of the period. Specifically, we folded the light curve on the best fit period using 16 phase bins, and determined the mean mag and standard error for each bin. We then computed the ratio between the range (i.e., the maximum bin value $-$ the minimum bin value) and the mean standard error across all bins. The cut we then chose, based on visual inspection of the light curves, is a minimum ratio of nine.

For the final two confidence cuts, we check the consistency of the period and the phase of the detected period by dividing the data into subsets. We did not require that every subset reproduce the values observed from the complete dataset, as physical changes in a system can result in significant changes in the detectability of the period. Nevertheless, we find it useful to require some degree of consistency in period and phase.

To explore the consistency of the period, we analyzed a rolling periodogram made from 300 consecutive points at a time. We then stepped forward 50 points and repeated the exercise until we reached the end of the dataset. We rejected a source from the ``confident" category if fewer than a third of the rolling peaks had a peak within 5\% of the overall periodogram peak. 

To then determine the consistency of the phase, we phase-folded non-overlapping groups of 300 points, this time with ten bins to make the phase of the peak less influenced by noise. 
However, we used groups of 100 points for SMC X-1 because it has very few data points but values under ten days for both the established period and periodogram peak.
For this additional check of consistency, we cut at a mean absolute change of two phase bins for the phase-folded peak.

After visual inspection of the results of the "confident" and "non-confident" categories, we added four sources to 
the "confident" category (Sources \hyperref[fig:type3gallery2]{51}, \hyperref[fig:type5gallery1]{61}, \hyperref[fig:type5gallery1]{107}, and \hyperref[fig:type5gallery1]{111}), and removed Source \hyperref[fig:type3gallery3]{\#106} for the reasons given below. It is this modified list that we use in
Sect. \ref{sec:discussion}. 

\begin{itemize}
    \item Source \#51 only has a strong periodic signal during the bright parts of the light curve, which occur toward the beginning and end of the OGLE coverage. Because of the intermediate weakening, this source fails to pass the final consistency check. However, the time-resolved phase-folds show the signal is present throughout with a consistent but noisy peak phase.
    \item Source \#61 and Source \#107 (Sect. \ref{sec:longperiod}) have infrequent, sharp periodic flares. Because of these features, the confidence procedure is not representative of the qualitatively apparent strength of the $\sim$656 and $\sim$1179 day periods. However, for Source \#61, both the longer period and the peak of the 2$-$200d periodogram pass the rolling periodogram check if no spline-detrending is performed. 
    \item Source \#111 undergoes a clear transition mid-light curve that coincides with a drastic weakening of the periodic signal. However, this signal starts strong and remains apparent in visual inspection.
    \item Source \#106 has a relatively short light curve and therefore only two chunks to use for the final consistency check. Although the two have consistent peak phases, we are hesitant to keep this source in the confidence category considering the limited data and the inconsistent LS peaks for the two light curve subsets.
\end{itemize}

The confidence metrics described above ignore any previously established periodicities. However, one could modify our conclusions based on the consistency, or lack thereof, with previously suggested values. In this regard, we note the following findings from our comparison of the uniform and targeted searches. 
Of the 50 sources with established optical periods between 2 and 200 days, 29 produced a 2$-$200 day LS peak equal to the targeted best period. Although we require these ``targeted" LS peaks to be within 10\% of the established period, the median difference between the reported period and the peak we find is under a day. Furthermore, for the sources with different overall and targeted best periods, there is no clear bias to higher or lower periods for the overall LS peak.

There are other notable features of these 29 sources. First, while there are established periods for sources in each super-orbital type, the sources with agreement between the overall and target LS peaks are all Type 2, 3, and 4. Second, the few confident sources with disagreement have harmonics of the established period as the overall peak. Thus, it is again clear that close attention should be paid to harmonics.

For a final comparison between these search windows, we contrasted the results of the confidence procedure for the targeted and untargeted searches. While the overall search peak tends to yield higher phase-folded range/error than the targeted peak, there are instances where this ratio passes the first confidence cut for the targeted but not the overall value. As discussed in Sect. \ref{sec:xrayperiods}, the Lomb-Scargle periodogram does not simply optimize phase-folded range. At the second confidence cut, the targeted search values that differ from the overall peaks do not meet the confidence requirement. Nevertheless, many do pass the peak phase consistency check and are thus still worthy of consideration as potential orbital periods.

\subsection{Periods of two to ten days}
\label{sec:short_periods}
We identify several sources with best periods between two and ten days, even though this range is unexpected in BeXRBs for both non-radial pulsations and orbital periods \citep[e.g.,][]{Rivinius2003,Bird2012}. 
One such source is SMC X-1, which is the only known Roche-lobe overflow HMXB in the SMC. Orbital periods under ten days are common for this class of HMXB \citep[e.g.,][]{Reig2011}.
Its best period is about half of its established orbital period of 3.89 days, probably because the orbital period yields a double-peaked phase-folded profile.
Although it is likely that the other sources' peaks largely result from aliasing or beating, it is worth noting a few characteristics of these situations.

First, we follow the confidence procedure outlined in Sect. \ref{sec:confidence} for sources for which the most prominent 2$-$200 day peak is under ten days. Although 26 sources meet this initial criterion, only 16 clear the FAL cutoff. Six of these sources exceed a ratio of nine between range and error. Of these sources, three have measured spin periods, and perhaps importantly, two (Sources \hyperref[fig:type6gallery1]{46} and \hyperref[fig:type6gallery1]{56}) are at the high end of both spin and previously reported optical periods, while the third is SMC X-1.

Five systems then pass the rolling periodogram consistency check and these rolling periodograms exhibit notable changes that could inform the interpretation of these systems. For instance, SXP 323 undergoes ``period drift" from $\sim$2.44 to $\sim$2.45 in a path similar to the drift of its $\sim$0.708 day period reported in \citet{Bird2012} and corroborated by our 0.1$-$2 day rolling periodogram. On the other hand, the shape of the overall downward drift of Source \hyperref[fig:type6gallery2]{\#110} resembles an inversion of the increasing drift for its $\sim$0.553 day period. Unlike these drifters, Source \hyperref[fig:type6gallery2]{\#76} exhibits a discrete period jump mid-light curve from 2.08 and 0.66 days to 2.17 and 0.68 days. Potentially explaining the behavior of SXP 323 and Source \#76 is the fact that the higher value in each situation is consistent with the beat period of the sub-2 day potential non-radial pulsation and one day (which is a factor of the ground-based sampling).

Two sources (Source \hyperref[fig:type6gallery1]{\#1}, which is SMC X-1, and Source \#76) join the confident category following the peak phase consistency check. If we compare the other confident sources and these two---or all aforementioned six meeting the range cut---we can note some differences akin to those explored in \citet{Bird2012}. Namely, using 16 phase bins, the minimum and maximum magnitudes for the short periods are 7$-$9 bins apart while over half of the longer-period confident sources are more asymmetrical than that. 
Still, we note that a few Galactic BeXRBs exhibit periods this short \citep[][]{Neumann2023}.
Thus, although the possibility is worth pursuing---particularly for potential Roche-lobe overflow systems---these short periods may not be orbital and instead may be beating between sampling and shorter-period pulsations.

\subsection{Periods of 0.1 to two days: Non-radial pulsations} 
\label{sec:NRPs}
Using both the detrended and original light curves, we searched for periods in the likely range for non-radial pulsations (0.1$-$2 days). Correlations between non-radial pulsations and other system quantities may elucidate the role of these pulsations in the formation and dissipation of the Be disk and the subsequent optical super-orbital behavior. 
However, the sampling of the majority of the OGLE data is not ideal for study of periods in this range. 
In particular, aliasing associated with once-per-night sampling is likely to be prevalent. To conservatively account for peaks due to sampling, we masked any power within 1/25th of each harmonic of a day (i.e., within 4\% of 1 day, 4\% of 0.5 days, etc.). 

Using this approach, according to the same FAL cutoff of 0.001, the majority of sources are then considered ``aperiodic" in this regime. In particular, only ten of the original light curves have significant potential non-radial pulsations. Six of these ten systems are precisely the ones discussed above in Sect. \ref{sec:short_periods}. Namely, their 2$-$200 day peaks are below ten days and yield phase-folded profiles with high enough range/error to pass the first confidence check (Sect. \ref{sec:confidence}). Furthermore, the Type 4 peaks are all above a day, and the Type 5 ones, with the exception of SMC X-1, are below.

With detrending, the number of sources with significant periodic signals increases to 14 and these sources include at least one system from each super-orbital type. We encourage a more rigorous cleaning approach and investigation into the relation between different periodicity peaks. Such an analysis could focus on the OGLE high-cadence data.

\subsection{Super-orbital periodicity}
\label{sec:super_periodicity}
As the final OGLE periodicity search, we checked for super-orbital periodicities. 
Inspired by the stability of the 421-day super-orbital oscillation in LMC BeXRB A0538-66 \citep[][]{Alcock2001}, \cite{Rajoelimanana2011} presented a linear correlation between orbital and super-orbital (quasi-)periodicities for a sample of 20 SMC BeXRBs.
More recently, \citet{Martin2023} proposed that precession of the Be star spin axis can cause BeXRB super-orbital periodicities, especially for systems with short orbital periods. 
However, we find that these SMC systems do not have true super-orbital periodicities when more data is incorporated. Instead, these sources are sometimes locally periodic (at least within the possible precision) with values that are not stable over longer baselines. We discuss one such source in Sect. \ref{sec:sxp6.85}.
In other cases, the approximate time scale is fairly stable but never exactly consistent. Source \hyperref[fig:type3gallery2]{\#69} is the only source with a potentially consistent super-orbital periodicity across our baseline of $\sim$8000 days, but it only shows three well-defined dips in this time. 

In the absence of clear super-orbital periodicities, we do not report values for super-orbital periods, and instead encourage future efforts to probe the specifics of these changing quasi-periodicities and the possibility to use some representative time scale in searches for correlations.

\begin{figure}
    \centering
    \includegraphics[width=\columnwidth]{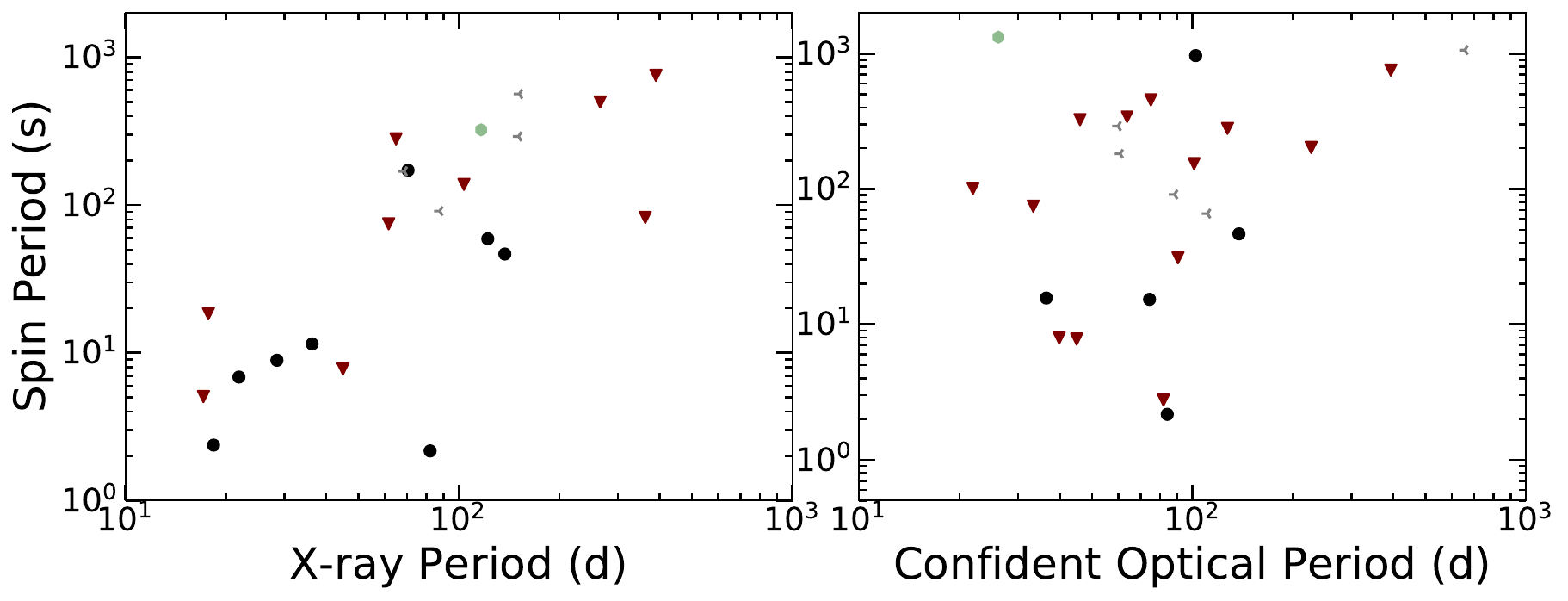}
    \caption{Corbet diagram with markers indicating super-orbital type. \textit{Left:} Spin period versus orbital period as proposed using X-ray data. \textit{Right:} Spin period versus best optical period for the 24 sources with both measured spins and confident periodicities (Sect. \ref{sec:confidence}). We exclude non-BeXRB SMC X-1, which has the shortest spin and orbital periods. 
    Of the eight sources present in both panels, three (Sources \hyperref[fig:type3gallery1]{23}, \hyperref[fig:type3gallery2]{41}, and \hyperref[fig:type5gallery1]{42}) have significant differences between the two potential orbital periods.}
    \label{fig:corbet}
\end{figure}
\begin{figure}
    \centering
    \includegraphics[width=\columnwidth]{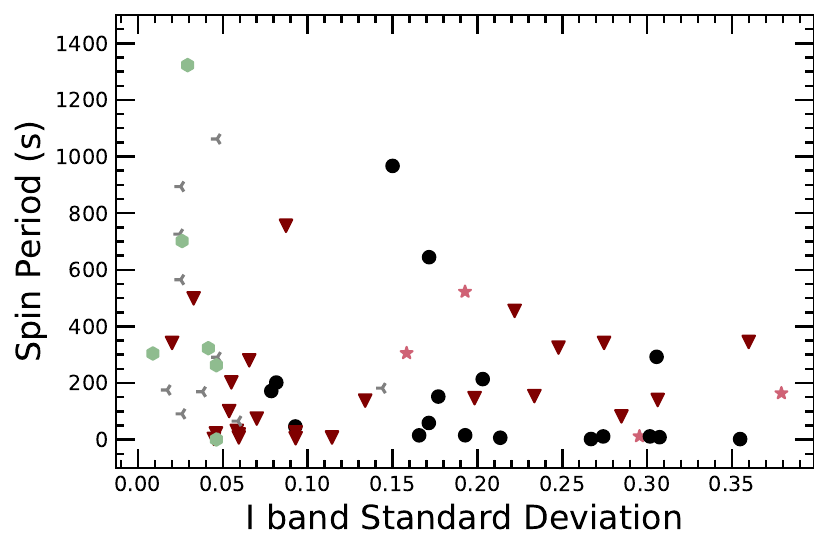}
    \caption{Neutron star spin period versus I-band standard deviation with sources marked by type. While sources with low optical variability have a range of spin periods, the most variable sources are only found with lower spin periods.}
    \label{fig:spinstdev}
\end{figure}

\subsection{Neutron star spin periods}
\label{sec:NS}
Correlations involving the NS spin periods help isolate the role of the NS star in the observed optical properties. Here we discuss two such relations: the well-known Corbet relationship between the orbital and spin periods, and the relationship between the spin period and the level of optical variability.

We use the periods cataloged in \citetalias{Haberl2016} for 
57 of the 63 systems in our sample with measured spin periods. In addition, the spin periods of SXP 15.6, SXP 164, and SXP 305 have been identified since then and we include recently discovered systems SXP 146 \citep{Kennea2020}, SXP 182 \citep{Coe2023,Gaudin2024}, and SXP 342B \citep{Maitra2023}.
The remaining systems in our sample are also X-ray sources with early-type optical counterparts but they lack pulsation measurements, either because the compact object is not an NS or the system has not yet been observed in X-ray outburst, at which point the spin period becomes more easily measurable. 
There is also some chance that these systems are misidentified as HMXBs.
Additionally, we note that spin periods can evolve significantly, so future work should account for updated values and potential correlations with period derivatives.

This spin period evolution likely gives rise to the Corbet relation.
Using seven systems, \cite{Corbet1984} identified a correlation for BeXRBs between the NS orbital and spin periods.  
The approximate correlation for these systems emerges from magnetospheric accretion regulating the NS spin period \citep[e.g.,][]{Reig2011}. In particular, rapidly spinning NSs lose angular momentum when their super-Keplerian magnetic field blocks accreting material from passing the magnetospheric boundary. On the other hand, for long spin period objects, this accreting plasma can reach the NS surface and provide angular momentum for spin-up. This dual regulation then relates to the system orbital period because orbital period affects mass accretion rate from the Be star toward the NS star, and mass accretion rate, in turn, sets the equilibrium period at which spin period can remain constant.

Subsequent studies of BeXRBs have added systems to the Corbet diagram \citep[e.g.,][]{Townsend2011}. In Fig. \ref{fig:corbet}, we plot spin period versus optical period for sources with measured spin periods and confident optical periodicities (see Sect. \ref{sec:confidence} and Table \ref{t.optical}). For comparison, we also show spin period versus X-ray-measured orbital period, which yields a tighter correlation. 
There are two Type 5 sources with both a spin period and a confident optical period but each one is unique in the sample: SXP 1323, which is consistent with a supernova remnant and spinning up by $\sim$30 seconds per year \citep[][]{Carpano2017,Gvaramadze2019,Mereminskiy2022}, and SMC X-1, which is the only known Roche-lobe overflow HMXB in the SMC.

In Fig. \ref{fig:spinstdev}, we show NS spin period versus the standard deviation of the corresponding I mag light curve. We find that sources with higher optical variability appear to have a lower maximum spin period. This relation could be an extension of the idea of persistent sources with low variability also having higher periodicities \citep[e.g.,][]{Reig1999}. 
\citet{Coe2005} presented a similar relation between spin period and variability and suggested an inverse power law relation between the quantities. However, with our larger sample, it becomes clear that the lower-variability systems can exist at a range of spin periods, even if there is some bias to higher spins.
Furthermore, \citetalias{Haberl2016} presented a loose anti-correlation between spin period and both X-ray flux and X-ray variability. Thus, on average, short time scale systems tend to be more variable in both the optical and X-ray regimes.

\section{Interesting sources}
\label{sec:interesting}
In this section we highlight the behavior of a few interesting sources. Although these sources are outliers in certain respects, the noted behavior is present in less extreme ways in other systems.

\subsection{Sources with bright \swift data}
\label{sec:swift}
Archival \swift data were available for six sources that underwent bright, well-monitored X-ray outbursts: Sources \hyperref[fig:type5gallery1]{5}, \hyperref[fig:type3gallery1]{6}, \hyperref[fig:type2gallery1]{7}, \hyperref[fig:type3gallery1]{8}, \hyperref[fig:type6gallery1]{44}, and \hyperref[fig:type3gallery2]{58}. Here, we compare these X-ray light curves to the behavior observed by OGLE.

The three systems with definite type II outbursts (SXP 4.78, SXP 5.05, and SXP 7.78) have low spin periods and relatively low optical variability, with the highest-amplitude feature coinciding with the X-ray outburst. All three are classified as Type 3 sources and have negative base numbers because of this flaring behavior.
\citet{Brown2019} proposed that SXP 5.05 \citep[Source \#6;][]{Coe2014} has orthogonal Be disk and orbital planes. The system underwent a type II outburst in 2013 \citep{Coe2015}. The X-ray peak coincides with the most significant brightness increase in the available coverage but the optical peak precedes the X-ray by tens of days.

The outbursts of SXP 6.85 (Sect. \ref{sec:sxp6.85}, Fig. \ref{fig:sxp6.85}) are less dramatic in both wavelength regimes. There are three X-ray outbursts that occur during optical peaks, but two are $\sim$200 days apart within the same optical peak (or, alternatively, they may represent one double-peaked X-ray outburst).

On the other extreme of spin period, SXP 756 (Source \#58, Fig. \ref{fig:longperiod}), exhibits bright type I X-ray outbursts that coincide with the periodic optical peaks. We further discuss this source in Sect. \ref{sec:longperiod}.

The behavior of SXP 304 (Source \hyperref[fig:type6gallery1]{\#44}) is the most ambiguous. The optical variability is relatively low, including during X-ray brightening events. The two observed outbursts, separated by $\sim$662 days, may be type I outbursts \citep[][]{Kennea2018}, but orbital periods proposed using the OGLE data have been shorter (344d and 520d), albeit with low confidence.

\subsection{Source \#18, Source \#47, and Source \#111}
\label{sec:bwb}
Only one source (Source \hyperref[fig:type3gallery1]{\#18}, which is SXP 25.5) is strictly ``bluer when brighter." Because the Be disk is cooler than the Be star, an expanding disk corresponds to a redder and brighter system unless the disk obscures the star along the line of sight, in which case a growing disk yields a redder and fainter system \citep[e.g.,][]{Hirata1982,Rajoelimanana2011}.
Thus, this color-magnitude relation suggests that the system hosts the only Be disk in our sample that is consistently edge-on. 

However, Source \hyperref[fig:type3gallery2]{\#47} and Source \hyperref[fig:type5gallery2]{\#111} each undergo super-orbital transitions at the same time as switches from redder-when-brighter to bluer-when-brighter behavior. The $\sim$0.5 mag brightening event of Source \#47 corresponds to a clear decrease in the amplitude of its orbital modulation. This behavior would be surprising if not for the source's simultaneous switch to being bluer-when-brighter.
In fact, it may be that the super-orbital variability is always bluer-when-brighter, with the earlier redder-when-brighter CMD just reflecting color changes related to the orbital time scale. 
Regardless, because of the emergence of a secondary peak in the phase-folded profile around this time, in addition to brightening in the X-ray, \citet{Coe2008} proposed a misalignment in this system.
This transition does seem consistent with an edge-on Be disk with a misaligned orbiting NS, but perhaps the disk precesses, making it only edge-on when we observe the bluer-when-brighter behavior. General disk growth would obscure the star, making it redder and fainter, but if the NS does not pass periastron along the line of sight, it may grow the disk without further obscuring the star.

In the case of Source \#111, the transition to bluer-when-brighter behavior coincides with a drastic change in orbital and super-orbital behavior, but at comparable magnitudes. \citet{Coe2021} explains the transition with the model of Be disk loss followed by 
regrowth, at which point the source returns to being redder-when-brighter. Even with this return, the qualitative change in the light curve persists. 

\subsection{Source \#7}
\label{sec:sxp6.85}
\begin{figure*}
    \centering
    \includegraphics[width=2.0\columnwidth]{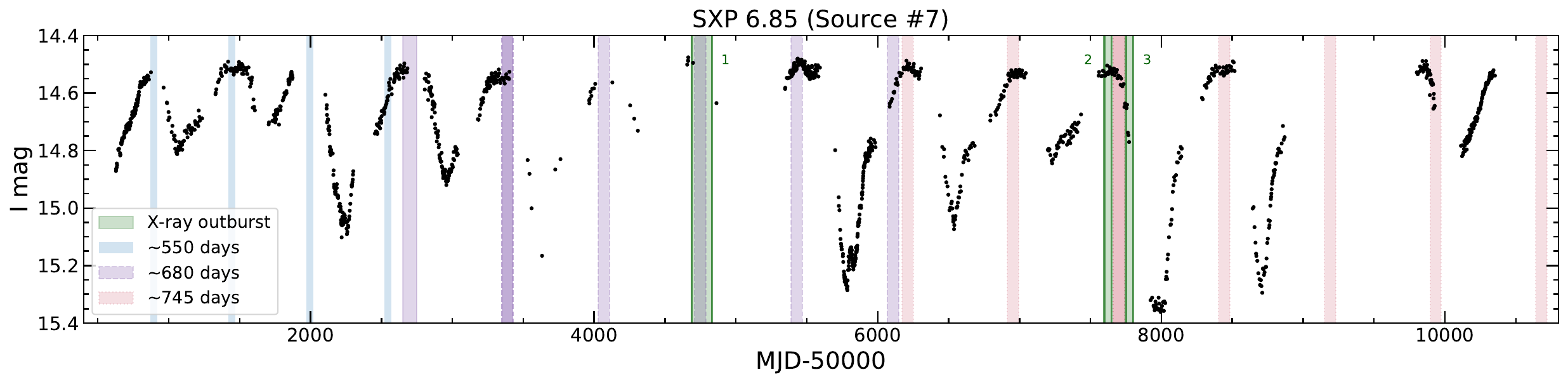}
    \caption{I-band light curve for Source \#7 (SXP 6.85). Three epochs with approximately stable super-orbital periodicities are shown; the third periodicity seems to fail with the most recent peaks but it is difficult to tell the next value without more data. Bright X-ray outbursts observed by \swift are also highlighted and numbered, with two of them coinciding with optical peaks.}
    \label{fig:sxp6.85}
\end{figure*}

There are multiple notable features of Source \hyperref[fig:type2gallery1]{\#7}, which is SXP 6.85 \citep{2003ATel..163....1C}.
More than any other source, SXP 6.85 resembles a flipped Type 1. However, the maximum, while stable, is more rounded than the Type 1 faint bases. As shown in Fig.~\ref{fig:sxp6.85}, the dip amplitudes are variable, with later times in the OGLE coverage including fainter dips. These super-orbital dips are not strictly periodic, but seem locally periodic in three epochs with best periods of $\sim$550, $\sim$680, and $\sim$745 days, respectively. These discrete, locally periodic epochs are similar to those found in LXP 69.5, an LMC BeXRB  \citep{Treiber2021}. In each system, the local periodicities are observed to be increasing, but have only been observed for three epochs. 

\cite{Townsend2011} noted the presence of quasi-periodic super-orbital behavior, with a large X-ray outburst coinciding with an optical peak.
Three observed X-ray outbursts from this source are highlighted in Fig.~\ref{fig:sxp6.85}. As \cite{McGowan2008} pointed out, the X-ray outbursts are not always correlated with optical peaks.

The shorter-term behavior of SXP 6.85 is also interesting. Using X-ray data, \citet{Townsend2011} found an orbital periodicity of 21.9$\pm$0.1 days. However, \cite{2015ATel.7498....1S} and \cite{McGowan2008} reported an optical periodicity of 24.8 days. We also find that value to be stronger than the X-ray period. However, as noted in the previous optical studies of the system, the periodogram is noisy and the overall peak is above 100 days ($\sim$109 days in our search).

\subsection{Source \#84}
\begin{figure}
    \centering
    \includegraphics[width=\columnwidth]{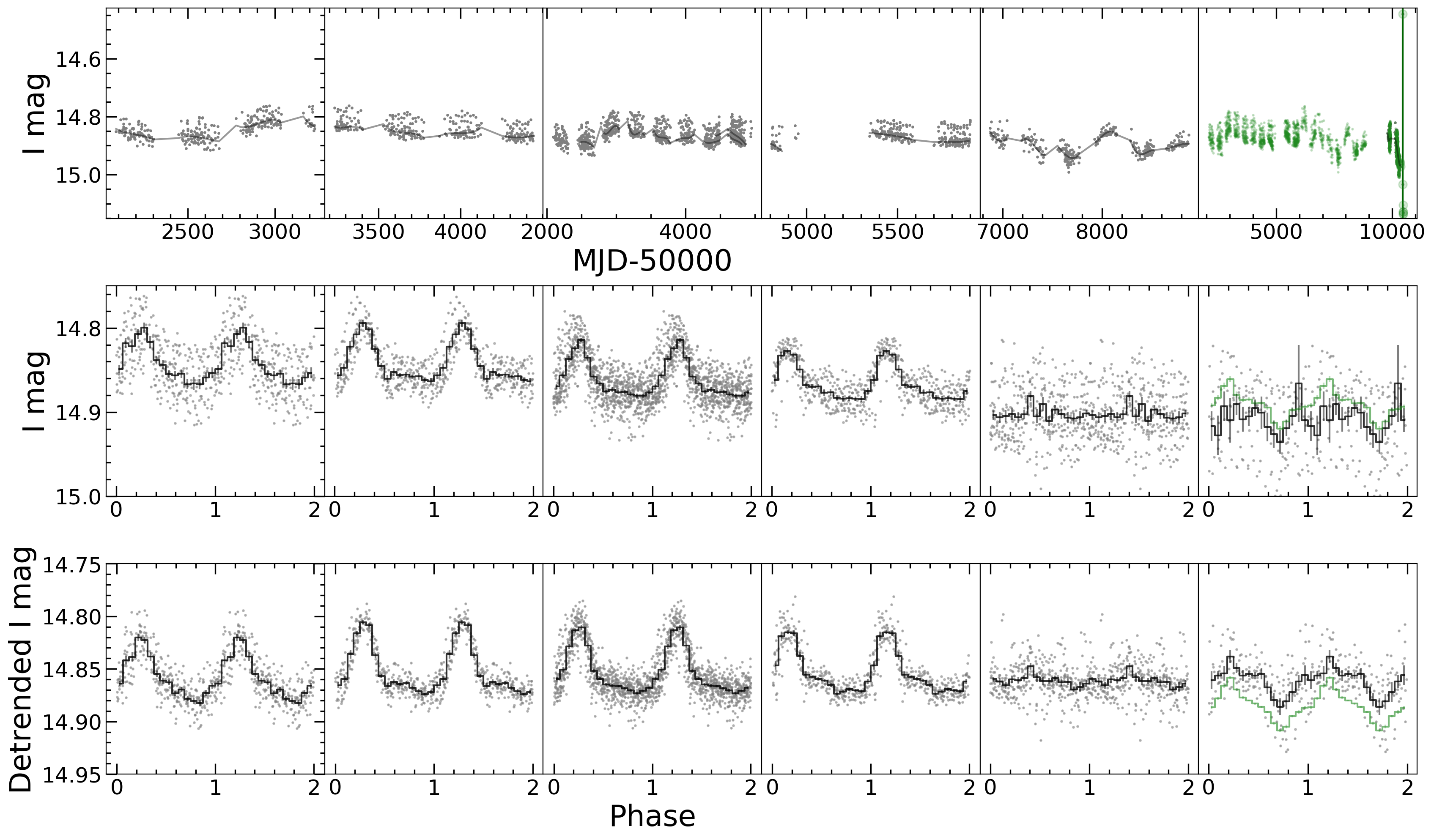}
    \caption{Light curve and evolving phase-folded profile of Source \#84. We divided the light curve such that the first five pieces include approximately 300 points. The gray curve is the spline used for detrending the light curve. The phase-folded profiles use the original data in the middle row and the detrended data on the bottom. 
    In green, we also plot the original high-cadence, unbinned data, with the final few points enlarged for emphasis. The vertical line marks the first observation of the May 2024 X-ray outburst \citep{Yang2024ATel}.
    Although the period (17.55 days) for this source is in the confident category (Sect. \ref{sec:confidence}), the signal seemingly disappears in the second to last piece of the light curve and then shows signs of reappearing more recently.}
    \label{fig:disappearing}
\end{figure}

In Fig. \ref{fig:disappearing}, we show the light curve of Source \hyperref[fig:type5gallery2]{\#84}, a Type 4 source without a measured spin period or established optical period. This binary is suspected to be a white dwarf BeXRB because of its X-ray properties during a May 2024 outburst
\citep[][]{Yang2024ATel,Kennea2024ATel,Jaisawal2024ATel,Marino2024,Gaudin2024WD}.

As we noted in \cite{Treiber2024},
we are confident in the periodicity of 17.55 days for this source. However, as shown in Fig. \ref{fig:disappearing}, when we break up the light curve into six sections, for example, the strong periodic signal seemingly disappears in the second-to-last chunk, whether or not we detrend the data. 
There is some visual evidence that the signal does not entirely disappear, but it clearly weakens. The periodicity then returns, but may still maintain a weaker phase-folded amplitude.
While similar disappearance is apparent in a few other sources (e.g., Source \hyperref[fig:type3gallery2]{\#51} and Source \hyperref[fig:type2gallery2]{\#95}), it usually coincides with a drastic brightness decrease, which is not the case here.

We also see notable behavior around the time of the X-ray outburst: the day that the outburst was first observed, the source drastically brightened beyond 14.5 mag, but within a couple days, it became the faintest it has ever been in the OGLE coverage.

\begin{figure*}
    \centering
    \includegraphics[width=2\columnwidth]{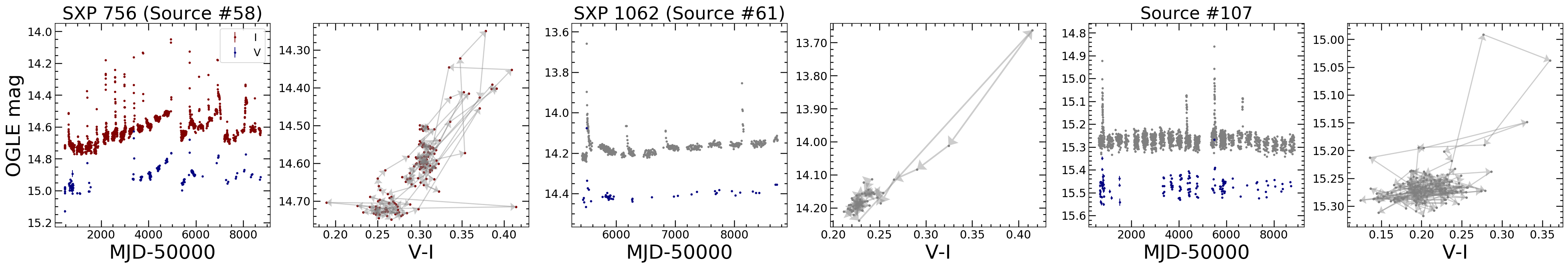}
    \caption{Light curves and CMDs of Sources 58, 61, and 107.}
    \label{fig:longperiod}
\end{figure*}

\subsection{Source \#58, Source \#61, and Source \#107}
\label{sec:longperiod}
Sources \hyperref[fig:type3gallery2]{58} and \hyperref[fig:type5gallery1]{61} (SXP 756 and SXP 1062) are two of the longest time scale systems and are also notable due to their extensive similarity. In addition to their long spin periods, these two sources have light curves with variability dominated by high-amplitude ($\lesssim$0.5 mag) brightening events coinciding with their 393- and 656-day orbital periods (Fig. \ref{fig:longperiod}). 

Source \hyperref[fig:type5gallery2]{\#107} has a similar light curve morphology but with flares separated by 1179 days. No spin period has been measured for this source, but one would expect it to also be one of the longest time scale sources with a spin of $\gtrsim$1000 seconds.
Furthermore, these sources are nearby in color-magnitude space (Fig. \ref{fig:allcm}), being among the reddest and brightest sources in the sample.
For Sources 58 and 61, the maximum H$\alpha$ equivalent width (EW(H$\alpha$)) measurements cataloged in \citetalias{Haberl2016} are $-$23.7 and $-$26.6, so both exhibit stronger H$\alpha$ emission than the median for the sample.

\section{Discussion}
\label{sec:discussion}
\begin{figure}
    \centering
    \includegraphics[width=\columnwidth]{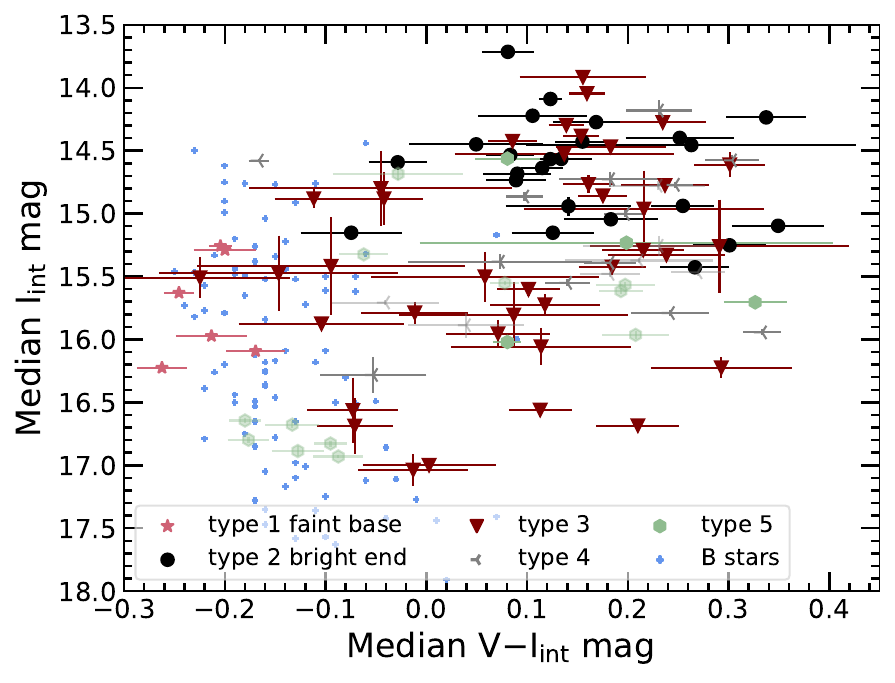}
    \caption{Color-magnitude diagram for the bases of Types 1 and 2 and for entire Type 3, 4, and 5 light curves. 
    Light blue points indicate SMC B stars in the \citet{Kourniotis2014} sample. For sources in our sample, error bars show one standard deviation in each axis for the given full light curve or base.
    Translucent points indicate that the corresponding source required manual calibration of OGLE epochs, entailing an additional source of uncertainty in both axes.}
    \label{fig:basec}
\end{figure}
\begin{figure}
    \centering
    \includegraphics[width=\columnwidth]{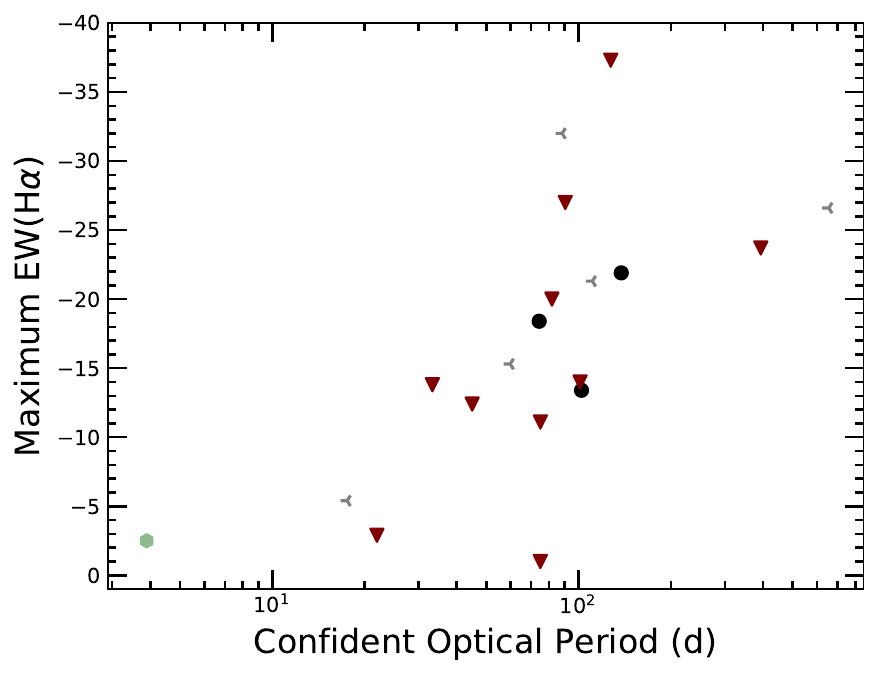}
    \caption{Maximum EW(H$\alpha$) versus the best optical period for the confident sources with EW measurements cataloged in \citetalias{Haberl2016} or for SXP 182 and SXP 342B in \cite{Gaudin2024} and \cite{Maitra2023}.}
    \label{fig:EW}
\end{figure}
We have presented a novel, quantitative approach to classifying the super-orbital variability of HMXBs. We explored the periodicities and color-magnitude behavior of the SMC systems. Here we investigate potential physical explanations for our five super-orbital types in the hope of better understanding HMXB orbital architectures, evolution, and how the NS (or other compact object) affects its counterpart.

\subsection{Disk changes and super-orbital variability}
\label{sec:type_explain}
We can explain the highly variable super-orbital behavior of Types 1, 2, and 3 (see Sect. \ref{sec:tax}) with a model of the relatively red Be disk growing and depleting. Across all BeXRBs, we expect the decretion disk around the Be star to be redder than the Be star \citep[e.g.,][]{Rajoelimanana2011}. 
For the remainder of the paper, we refer to the Be decretion disk as simply the ``disk," which should not be confused with an accretion disk that might appear around the NS. 
For most systems, as the disk expands, we should see the source become both redder and brighter. Concurrent photometric and spectroscopic observations of Be stars also support the notion that outbursts coincide with disk growth \citep[][]{Labadie-Bartz2017}. As mentioned in Sect. \ref{sec:bwb}, only one source in our sample shows consistent bluer-when-brighter behavior.  
However, Sources \hyperref[fig:type3gallery2]{47} and \hyperref[fig:type5gallery2]{111} also present this behavior, but only for $\sim$3000 days and $\sim$1000 days, respectively.
This low bluer-when-brighter incidence is consistent with a line-of-sight explanation, in which the rare case of an edge-on system will have the growing disk obscure the bright blue Be star \citep[e.g.,][]{Hirata1982}. Thus, in contrast to most other systems,
an edge-on source will become fainter \rm and redder as the disk grows.

The faint blue baseline and the redder flares of the Type 1 sources suggest that these sources have a small or nonexistent baseline disk that expands and fades again with some regularity. This interpretation is supported by the fact that Type 1 sources in their low baseline occupy the same part of the CMD as the \citet{Kourniotis2014} SMC B stars, which lack decretion disks entirely. 
They do not, however, span the full color-magnitude range that the B stars do, suggesting that a system's spectral type may place constraints on its super-orbital type.
Furthermore, as we would predict for a Be star with little to no disk material to accrete onto the star, \citet{Monageng2020} found that Source \hyperref[fig:type1gallery]{\#103} has relatively low X-ray flux that does not necessarily coincide with its Type 1 flares. Similarly, they found H$\alpha$ in emission
during a flare but it had previously only been observed in absorption \citepalias{Haberl2016}.

By contrast, we suggest that the Type 2 bright, red baseline corresponds to an approximate maximum disk size for a given source and the dips involve a shrinking disk followed by replenishment back to the maximum. That is, we suggest that these dipping light curves correspond to the depletion of the disk from some maximal value.  

The majority of Type 2 dips are still redder and brighter than Type 1 faint bases, suggesting the disks do not fully deplete. 
However, two Type 2 sources become as blue, but still brighter, and one is firmly among the Type 1 faint bases in color-magnitude space. The I-band light curves of these Type 2 sources also have qualitative similarities to the Type 1 sources.
The Type 2 sources are generally a more diverse group than Type 1 sources, in terms of both super-orbital appearance and color-magnitude behavior. The greater range of behavior may be due to differences from source to source of the underlying B star as well as of the maximal extent of the disk and the specifics of the depletion events. 

Similarly, by comparing the bright ends of the Type 1 flares to the Type 2 bright baselines, we see that the Type 1 flares do not reach the kind of disk maxima that we see as the baselines for Type 2 sources. 
Instead, the bright ends of Type 1 flares remain generally bluer and fainter than their Type 2 bright counterparts. 
Furthermore, Type 1 source SXP 164 seems to be transitioning into a later type following a brightening event that surpassed its flares' maxima.
The most consistent overlap in color-magnitude space is the bright end of Type 1 sources and the faint of Type 2. This is as one would expect if systems in these states had noticeable but non-maximal disks.

Type 3 sources have strong coherent variability, but lack a clear base either at minimum or maximum light to which the sources return, although sources closer to the Type 1 and Type 2 boundaries show potential baselines that may be confirmed with more coverage. In general, though, Type 3 systems appear to be intermediate sources that do not reach the extrema of completely depleted or maximal disks. As predicted by this explanation, Type 3 sources are generally not as blue as Type 1 sources nor as red as Type 2 sources.

Because we expect different disk maxima for different sources, it is unsurprising that when Type 3 sources are bright, they are consistent in color-magnitude space with some of the well-defined Type 2 bright maxima. For some of these systems, it is possible that their bright end is their disk maximum but they do not linger there sufficiently long for there to be a clear brightness maximum in the OGLE data. Likewise, there is significant overlap at the faint end for Types 2 and 3, although 16 Type 3 sources (41\%) become fainter than all Type 2 sources. 

Many Type 3 sources are comparable at their faint blue end to the faint baseline of the Type 1 sources, as would be expected. 
However, we note that eight Type 3 sources become fainter than all the Type 1 faint bases, although they remain redder even then and thus not clearly consistent with the B stars in Fig. \ref{fig:basec}, presumably because of the presence of the Be disk.
It is likely that the underlying B stars are somewhat later spectral types than the Type 1 sources (e.g., without a disk, they are more consistent with the \cite{Kourniotis2014} stars at $\mathrm{\sim17^{th}}$ mag). In that case, they are more likely to appear as Type 3 than Type 1, because even a faint disk will noticeably increase the flux above the level of the star itself.
This distinction is also consistent with the Type 3 sources generally having lower levels of coherent variability, since greater variability has been observed in earlier Be star spectral types \citep[][]{Hubert1998,Rivinius2013,Labadie-Bartz2018}.

\subsection{Maximal disks}
\label{sec:max_disks}
Beyond spectral type, we can attribute some of the diversity within Type 2 sources to varying disk maxima, whereas the Type 1 faint bases are self-similar because those bases correspond to (nearly) bare, similar stars. As shown in Fig. \ref{fig:basec}, the Type 1 sources are consistent in color-magnitude space with a subset of the SMC B stars studied in \citet{Kourniotis2014}, unlike the vast majority of systems in our sample. 
The question, then, is whether we can identify effects that determine a given disk maximum.

Truncation is one clear effect that would yield different disk maxima in different systems. Tidal truncation by the NS yields more compact, denser Be disks in BeXRBs than in isolated Be stars by a factor of $\sim$2 \citep[e.g.,][]{Reig1997,Negueruela2001,Zamanov2001,Okazaki2001,Okazaki2002}. 
Applying the model presented in \citet{NegueruelaOkazaki2001}, \citet{Okazaki2001} conclude that eccentricity is the key variable for explaining the truncation radius of BeXRBs. 
They suggest that truncation is inefficient in a high-eccentricity system, so it should be more straightforward to observe type I outbursts in such a case. 
Six sources have eccentricities cataloged in \citet{Coe2015} and the two with the highest eccentricities are also the reddest, favoring this picture of inefficient truncation.

Although eccentricity may be the most helpful variable for predicting truncation radius, we also expect this effect to be more pronounced in systems with shorter orbital periods \citep[e.g.,][]{Reig2011}. 
Despite the significant scatter for similar orbital periods, there is a correlation between orbital period and maximum EW($\mathrm{H\alpha}$) \citep{Reig1997,Antoniou2009}. This relation provides convincing observational evidence of truncation since $\mathrm{H\alpha}$ emission is a good proxy for the size of the disks' emitting region \citep[e.g.,][]{Quirrenbach1997}. Potentially contributing to the scatter is the lack of sufficient coverage to identify a representative maximum EW.
In Fig. \ref{fig:EW}, we present our version of this relation using only the confident orbital periodicities.
Similarly, shorter periods should also generally cause bluer maximal disks.
We see some evidence of this effect when comparing potential orbital periods and color, with a suggestion that longer periods result in redder maxima. However, the current numbers and scatter prevent a definitive conclusion, especially in the absence of sufficient information about the spectral types of the underlying stars.

Consistent with this explanation are SXP 756 and 1062 (Sect. \ref{sec:longperiod}), which have long spin and orbital periods. 
Large-amplitude variations dominate their variability and coincide with the orbital periods. One has \swift observations of type I outbursts. Moreover, these two systems are particularly bright and red, further supporting the notion that they are not as truncated as other sources in the sample.

On the other hand, \citet{Okazaki2001} expect moderate eccentricity systems to be efficiently truncated and the most likely systems to exhibit type II outbursts. The four systems with \swift type II outbursts have spin periods in the low range of 4.78 to 7.78 seconds. Three  of these sources show their highest-amplitude variability during the X-ray outbursts and the fourth is SXP 6.85, discussed in Sect. \ref{sec:sxp6.85}. Only SXP 7.78 has a clear, confident optical orbital signal (of 44.91 days). These systems are consistent with the model predictions, but a full understanding of disk truncation will require closer observed constraints on eccentricity, inclination, and maximum EW(H$\alpha$).

\subsection{Orbital periodicity and super-orbital variability}
\label{sec:period_types}
Lending further support to this picture is the apparent orbital behavior of these systems. If this physical explanation is correct, we expect clearer orbital periodicity signals for Type 2 sources and, with some scatter, for redder sources, because the Be disk is more likely to be sufficiently extended to be affected by the NS during its periastron passage.  By contrast, we would expect low or no periodic signal corresponding to the NS orbital period for a nearly or entirely depleted disk.
Indeed, we find that none of the Type 1 systems have confident orbital periodicities (see Sect. \ref{sec:confidence} and Table \ref{t.optical}). Moreover, if we separately fold the Type 1 detrended bases and flares, we generally see stronger LS periodogram peaks and higher folded amplitudes during the flares.  Further exploration of this difference would be necessary to conclusively add confidence to the physical explanation.
On the other hand, we present confident periodicities for 25\% and 41\% of Type 2 and 3 sources, respectively. 

Furthermore, this connection between more extended disks (i.e., generally brighter and redder ones) and orbital period signal strength may extend to behavior within individual light curves. \citet{Rajoelimanana2011} identified a correlation between I magnitude and outburst amplitude for ten SMC sources and an anti-correlation only for one. 
While we often find insufficient evidence or periodic signal for this conclusion, Source \hyperref[fig:type2gallery1]{\#20} and Source \hyperref[fig:type2gallery2]{\#95} each show a clear relationship between brightness (or redness) and phase-folded amplitude. These two sources are also visually similar, with one brief dip of $\sim$0.4 mag in the OGLE light curve and orbital variations visible even on the super-orbital scale.

\subsection{Low coherence variables: Types 4 and 5}
\label{types_4_5}
Next, we turn to Type 4 and Type 5 systems, which are
characterized by low coherent variability and thus
high values of the stochastic variability metric. We argue that Type 4 sources are generally consistent with the processes described so far and can be explained as systems in which the disk has changed little during our observing window, perhaps because they are generally longer time scale systems. The Type 4 systems resemble Type 2 sources near maximum and may eventually show depletion events. On the other hand,
Type 5 sources may consist of a superposition of different kinds of systems, including:
(1) systems in which the disk is extremely stable on long time scales, (2) systems hosting a black hole, or other kinds of sources unexpected in the sample, or (3) mistakenly identified optical counterparts. 
Thus, from a physical point of view, many of the Type 5 systems may not be part of this sample at all.

Unlike Type 5 sources, which have the highest values of the stochastic variability metric, Type 4 sources still have some coherent long-term ($\gtrsim$ hundreds of days) variations. 
Exploration of these Type 4 sources reveals that the truncation story is not so straightforward. These 21 sources have some super-orbital coherent variability but with amplitudes comparable to the shorter-term, often orbital variations displayed by Type 2 and 3 systems, without the large super-orbital variations. Of Type 4 sources, 38\% are in our confident periodicity category. This finding does not seem to be a mere product of detrending difficulties for the more variable sources. Instead, despite the significant spread of Type 4 sources in color-magnitude space, we note that the median Type 4 color is redder than that of its Type 1, Type 2, and Type 3 counterparts. This suggests the presence of maximal disks in these systems. However, among the confident periodicities, there is no clear correlation between phase-folded amplitude and color. Furthermore, for the Type 4 sources with confident orbital periods, the systems with longer orbits are not redder and therefore do not necessarily host more extended disks than their shorter period counterparts. It may be that spectral type and inclination effects complicate the potential relationship between redness and disk size.

Type 4 sources may become Type 2 sources by showing depletion events in future coverage, and indeed they exhibit evidence of being longer time scale systems than their Type 2 counterparts. 
For the Type 2 and 4 sources with measured spin periods, the median and mean Type 2 spin periods are 47 and 166 seconds, whereas these values increase for Type 4 sources to 237 and 743 seconds, respectively. The difference in orbital periods is harder to discern, but the orbital periods established through X-ray data also share this time scale difference.  It seems plausible that the super-orbital time scales may also be longer in Type 4 systems.
We thus predict that some Type 4 sources will either lack depletion events entirely or show less frequent ones. Furthermore, this time scale connection is additional evidence of the effect of the orbital architecture on the super-orbital variability of the Be disk.

Finally, we motivate the need for the separation between Type 4 and 5 sources. Type 5 sources have little or no coherent super-orbital variability. Beyond that, though, we note differences in periodicities and type transition likelihoods between Types 4 and 5. First, while 43\% of our overall sample and 48\% of Type 4 sources have no measured spin period, this percentage jumps to 71\% for Type 5 sources. Furthermore, the Type 5 sources without spin periods have a mean I magnitude one magnitude fainter than those with spin periods. This subset of systems may thus be interlopers in our overall sample. Second, 67\% of the 21 Type 5 sources have no periodic signal above a false alarm level of 0.1\% in the 2$-$200 LS periodogram. Crucially, that percentage decreases to 33\% if we only consider the Type 5 sources with measured NS spin periods and again to 19\% sample wide.  

Moreover, the Type 5 super-orbital variability is low enough that these sources seem less likely to ``become" other types. As described in Sect. \ref{sec:tax}, 84\% of sources in Types 1--3 have at least 1000 consecutive days in their light curves that resemble Type 4 sources (i.e., that fall in the Type 4 region of Fig. \ref{fig:lbase_stdev}). On the other hand, only 54\% sources match the same criterion for resembling Type 5 sources. In addition, most Type 4 sources overlap with Type 2 bright ends in color-magnitude space (Fig. \ref{fig:basec}). Meanwhile, a cluster of six Type 5 sources are consistent with SMC B stars studied in \citet{Kourniotis2014}.

The question remains why a given source is a given type and how much the answer is a function of the observing window as opposed to something more fundamentally physical about the system, such as the spectral type. From Fig. \ref{fig:spinstdev}, we can see the role of spin period (and therefore according to \citet{Corbet1984}, orbital period as well). In particular, the most variable systems all have short spin periods, whereas many Type 4 and Type 5 sources host NSs with long spin periods. 

\subsection{Color-magnitude diagram loops}
\label{sec:loop_explain}
Further contributing to our view of the disk growth and depletion events are the loops that are primarily present in the Type 1 and 2 sources' individual CMDs. We describe these loops in Sect. \ref{sec:color} and show examples in Fig. \ref{fig:loop_examples}. 

\citet{deWit2006} modeled these loops with the following stages. First, as mass loss from the Be star turns on, an optically thick disk forms and expands, yielding the initial redder-when-brighter relation. 
The disk eventually ceases expansion and the rate of brightening diminishes and stalls completely. Then, when the outflow of material from the star decreases, the disk diminishes from the inside out, generating a ring-like structure with a hollow center. As a result, the blue light (from the inner disk) diminishes faster than the red light, so a decrease in overall flux is accompanied by even greater reddening. Eventually the disk dissipates altogether, and the flux and color return to their original values. This sequence gives rise to a clockwise loop through the CMD, as shown in Fig. \ref{fig:loop_examples}. This model can also account for systems that show this flattening at bright values without a full loop. 

Although \citet{deWit2006} focused on modeling systems that we would classify as Type 1, the explanation seems to work for loops arising from dips as well. Focusing on the I- and V-band light curves for Type 1 and Type 2 systems (Fig. \ref{fig:loop_examples}), we see the bluer flux decline faster than the redder flux from the Type 1 flare peak, likely because of the disproportionate clearing of the bluer optically thick inner disk. Likewise, the V-band dip precedes the I-band for Type 2 sources, whereas the two bands overlap during flux increase. Similar behavior is apparent in the LMC \citep[e.g.,][]{Haberl2023}.

In our overall sample, we notice these loops in 19 sources. However, largely due to data gaps in V-band, we do not regard our sample of loops as definitive. 
Only three of these looping sources (Sources \hyperref[fig:type3gallery1]{16}, \hyperref[fig:type3gallery1]{24}, and \hyperref[fig:type3gallery2]{50}) are neither Type 1 nor Type 2, and they have flares in their light curves that resemble those of Type 1 sources. Likewise, a few additional Type 3 sources have partial evidence for loops in their light curves and these sources' light curves also include more Type 1-like features than the typical Type 3 source.
All the observed loops are clockwise, which, according to \citet{deWit2006}, indicates that outflowing material (at $\sim$1 $\mathrm{km\,s^{-1}}$) drives the disk changes, as is expected to be the case for ``decretion" disks.  Thus the behavior of the loops provides further confirmation of the basic scenario, that is, the super-orbital variability is due to changes in the size, flux, and color of a Be decretion disk.

\section{Conclusions and future work} 
\label{sec:conclusion}
Using OGLE I- and V-band light curves of 111 systems, we investigated over two decades of optical behavior of the HMXBs in the SMC.
Here, we summarize our progress in classifying the long-term super-orbital behavior, searching for periodicities, understanding color evolution, and connecting these findings to each other and the cataloged X-ray behavior of these systems. We then list a number of suggestions for future work for observers and theorists alike.

Our main conclusions are as follows:
\begin{enumerate}
    \item Our two classifying metrics of base number and stochastic variability metric effectively reflect the qualitative differences between flaring, dipping, and intermediate sources and the level of coherent super-orbital variability, respectively. These quantities are useful discriminators as continuous variables, but for the purposes of this paper, we made cuts in each axis to produce five super-orbital types. We conclude that Types 1--3 can generally be interpreted with the model of the relatively red disk growing and depleting, with the cohesive group of Type 1 sources displaying a faint blue base consistent with B0 and B1 stars with intermittent disk growth events causing the observed flares. On the other hand, the return of the Type 2 sources to a consistent and bright level represents the systems' disk maxima, from which the sources show dips during partial depletion events. Within the available coverage, Type 3 sources do not reach these consistent disk minima or maxima, and some are redder and fainter than Type 1 bases, suggesting the underlying stars are a later type than the Type 1 stars, which is also consistent with their lower variability. However, BeXRBs with known spectral types are all earlier than B5 \citep{McBride2008}, so we must also consider explanations in which the underlying Type 3 star is significantly brighter and bluer than the star plus disk that we observe.
    Still, this large, diverse type would benefit from sub-classification. 
    Type 4 sources show lower levels of coherent variability, with their super-orbital variability no longer dominating the light curve. In their brightness and color evolution, many of these sources appear to be longer time scale Type 2 sources without dips and therefore may show depletion events in the future. Finally, Type 5 sources have the lowest levels of variability, and many represent fundamentally different sources.
    \item All but one of the sources become redder when brighter, although a couple of sources switch to being bluer when brighter following transitions that also lead to weaker orbital signals. 
    Such transitions may aid in the study of precessing and misaligned systems.
    Other complex features are common in the sources' color evolution, the most notable of which are the loops in the CMDs of many Type 1 and 2 sources and a few Type 3 light curves. These loops proceed clockwise in the CMD, which is behavior that has previously been modeled as an increase in the inner radius of the disk, as would be expected from a decretion disk with a falling mass injection rate.
    \item By relating super-orbital behavior to system periodicities, we can constrain how the NS affects the Be disk evolution. 
    In the search for orbital and other short-term optical periodicities, we found it particularly helpful to detrend the super-orbital variability with a spline from \texttt{wotan} \citep[][]{Hippke2019}. Primarily using Lomb-Scargle periodograms and informed by a comparison between X-ray and optical orbital signals, we ultimately found 33 confident potential orbital periodicities, five of which are for sources without proposed periodicities in the literature (Sources \hyperref[fig:type3gallery2]{64}, \hyperref[fig:type6gallery2]{76}, \hyperref[fig:type3gallery3]{88}, \hyperref[fig:type3gallery3]{97}, and \hyperref[fig:type5gallery2]{107}).
    We also refined previously established periodicities and more data or attention may confirm some of our lower-confidence periodicities.
    Even for confident sources, the periodic signal can abruptly disappear, often during faint parts of the light curve. More modest changes in phase-folded shape are common and may speak to how changes in the Be disk then alter the interaction with the NS at periastron.
    While we found no super-orbital periodicities or consistent correlations between these time scales, we note that the presence of confident orbital periodicity is significantly more likely for sources near disk maximum. 
    \item Spin period also contributes to the determination of super-orbital type for a given source. For instance, highly variable optical light curves only correspond to NSs with relatively low spin periods (which, by the Corbet relation, tend to also have short orbital periods). These systems also host denser, more truncated disks, which then have more modest values of H$\alpha$ emission despite their relative brightness. 
    At the other end of the time scale distribution are a few systems with the highest phase-folded amplitudes and repeated type I outbursts.
    These qualities illustrate the differential effect of the NS on its early-type host depending on the orbital architecture.
\end{enumerate}
Furthermore, while we focus on the population and only note the intricacies of a few sources (Sect. \ref{sec:interesting}), these questions would benefit from attention to unique sources or smaller groups of similar systems.

We suggest the following efforts in future work:
\begin{enumerate}
    \item There is increasing access to optical/NIR photometry of the SMC. 
    A clear next step would involve the higher-cadence OGLE light curves and perhaps TESS \citep[][]{Ricker2014,Reig2022},
    ASAS-SN \citep{Shappee2014}, and soon for the faintest objects in the sample, the Vera Rubin Observatory \citep{Ivezic2019}.
    The different cadences are helpful for searching for periodicities across different time scales and accounting for the beating effects present in lower-cadence OGLE data.
    Upcoming data could also inform our understanding of super-orbital types and transitions between them. 
    As a result, finer-grained taxonomies and corresponding explanations may become increasingly plausible and useful.
    \item While it is typical to assume all SMC stars are at a fixed distance, future work should account for recent radial-velocity constraints that suggest that the SMC hosts two star-forming systems separated by $\sim$5 kpc along the line of sight \citep{Murray2024}.
    \item A systematic study of the connections between orbital and super-orbital properties would better constrain how the NS affects the Be disk. Such a study would benefit from defining some super-orbital time scale in lieu of super-orbital periodicity. Beyond time scale correlations, specific connections between super-orbital features and phase-folded morphology could inform physical interpretations. For instance, double-peaked phase-folded profiles may indicate that the Be disk and orbital plane are misaligned such that the NS passes through the disk twice per orbit \citep[e.g.,][]{McGowan2007,Coe2009}. 
    \item Similarly, because of their potentially central role in the Be phenomenon, one could focus on non-radial pulsations and investigate whether their beat periods correlate with the time scales of drastic disk formation or depletion events.
    \item These time scale-focused studies would benefit from additional progress in methodologies to find periodicities and assign confidences.
    \item Another significant improvement to time scale studies and correlations would involve new and aggregate X-ray studies of spin period ranges and long-term X-ray behavior.
    \item A more detailed use of optical spectroscopy would clarify the physical picture behind the observed variability. For instance, the multi-year time scales of V/R variability in Be stars are consistent with the super-orbital variability discussed here. Since V/R variability seems to arise from one-armed oscillations in the disk \citep[][]{Okazaki1985,Okazaki1991}, the potential correlation would elucidate the physical explanation of the light curve variability. Furthermore, relationships between line and continuum emission could constrain disk truncation and density levels.
    \item Through these additional observational efforts and disk modeling, one could also potentially connect these OGLE data to spectral type, orbital architecture, warping, precessing, and disk parameters, including radius, density, viscosity, and temperature.
    \item Studies of HMXBs (and BeXRBs in particular) in the Milky Way and LMC could inform findings on the specific and population level. Similarly, a key line of evidence for the impact of the compact object would be a statistical comparison between the optical variability of isolated Be stars and Be stars in XRBs.
\end{enumerate}

The optical behavior of HMXBs offers valuable insights into the complex interactions between compact objects and their high-mass donor stars. Through our classification of super-orbital variability and its links to other system properties, we have highlighted promising avenues that will lead to a more comprehensive understanding of this diverse population.
\section{Data availability} 
Tables \ref{t.xray} and \ref{t.optical} are available in electronic form at the CDS via anonymous ftp to cdsarc.u-strasbg.fr (130.79.128.5) or via http://cdsweb.u-strasbg.fr/cgi-bin/qcat?J/A+A/.

OGLE-IV I band light-curves are available via the X-Ray variables OGLE Monitoring (XROM) system; \url{https://ogle.astrouw.edu.pl/ogle4/xrom/xrom.html}.
The data and code underlying this article will be shared on reasonable request to the corresponding author.

\begin{acknowledgements}
We thank the anonymous referee for their comments and suggestions that helped improve the clarity of the manuscript.
The OGLE project has received funding from the National Science Centre, Poland, grant MAESTRO 2014/14/A/ST9/00121 to AU.
HT acknowledges support by the National Science Foundation Graduate Research Fellowship Program under Grant DGE-2039656. 
Any opinions, findings, and conclusions or recommendations expressed in this material are those of the authors and do not necessarily reflect the views of the National Science Foundation.
HT is funded by the Dorrit Hoffleit Undergraduate Research Scholarship at Yale. 
HT and GV are supported by NASA Grant Number 80NSSC21K0213, in response to NICER cycle 2 Guest Observer Program. 
GV acknowledges support from the Hellenic Foundation for Research and Innovation (H.F.R.I.) through the project ASTRAPE (Project ID 7802). 
\end{acknowledgements}

%
%

\bibliographystyle{aa} 
\bibliography{general}
\appendix
\section{Light curve and color-magnitude gallery}
\label{sec:gallery}
\begin{figure*}
    \centering
    \includegraphics[width=2\columnwidth]{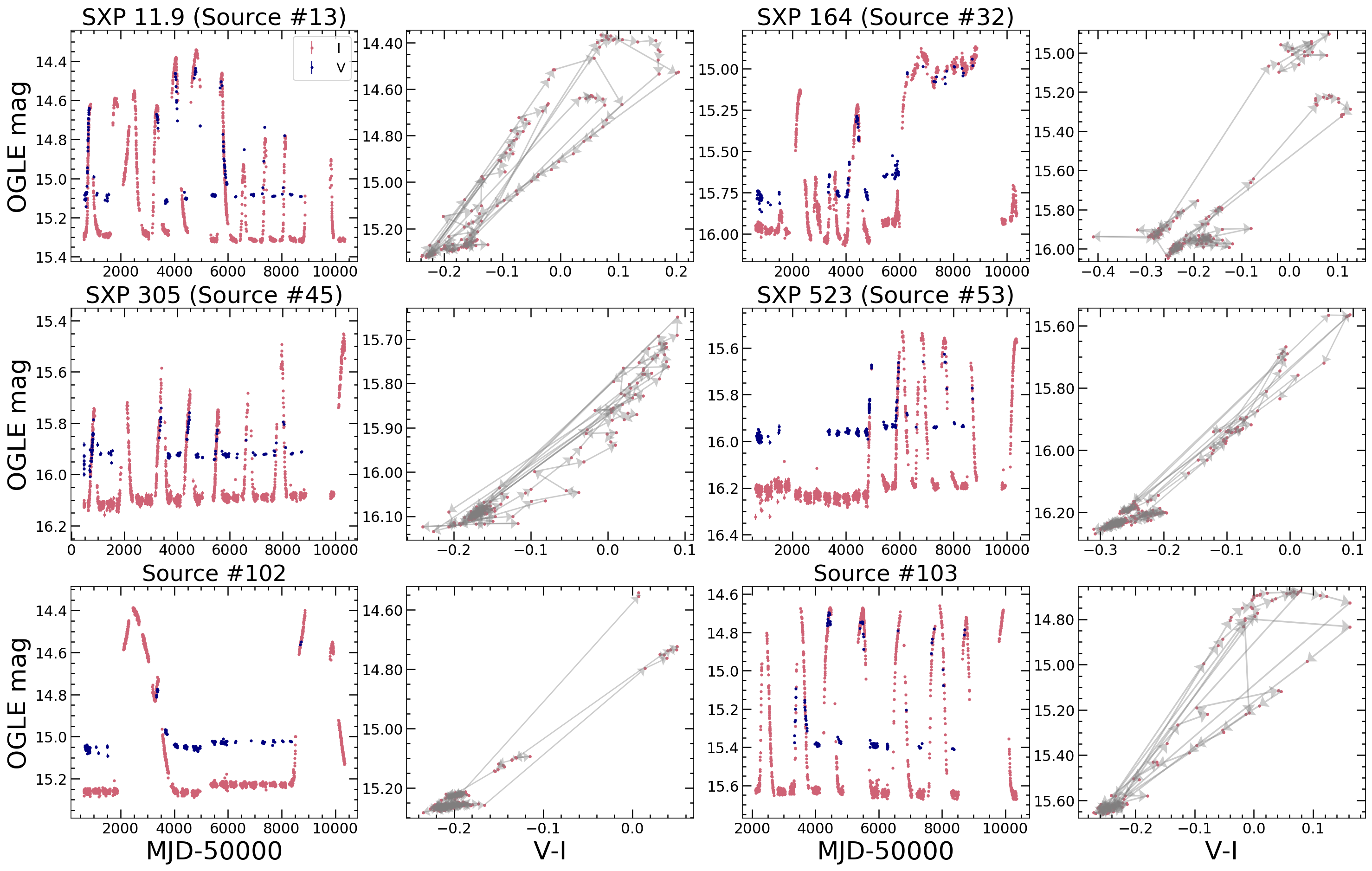}
    \caption{Light curves and color-magnitude diagrams for Type 1 sources.}
    \label{fig:type1gallery}
\end{figure*}

\begin{figure*}
    \centering
    \includegraphics[width=2\columnwidth]{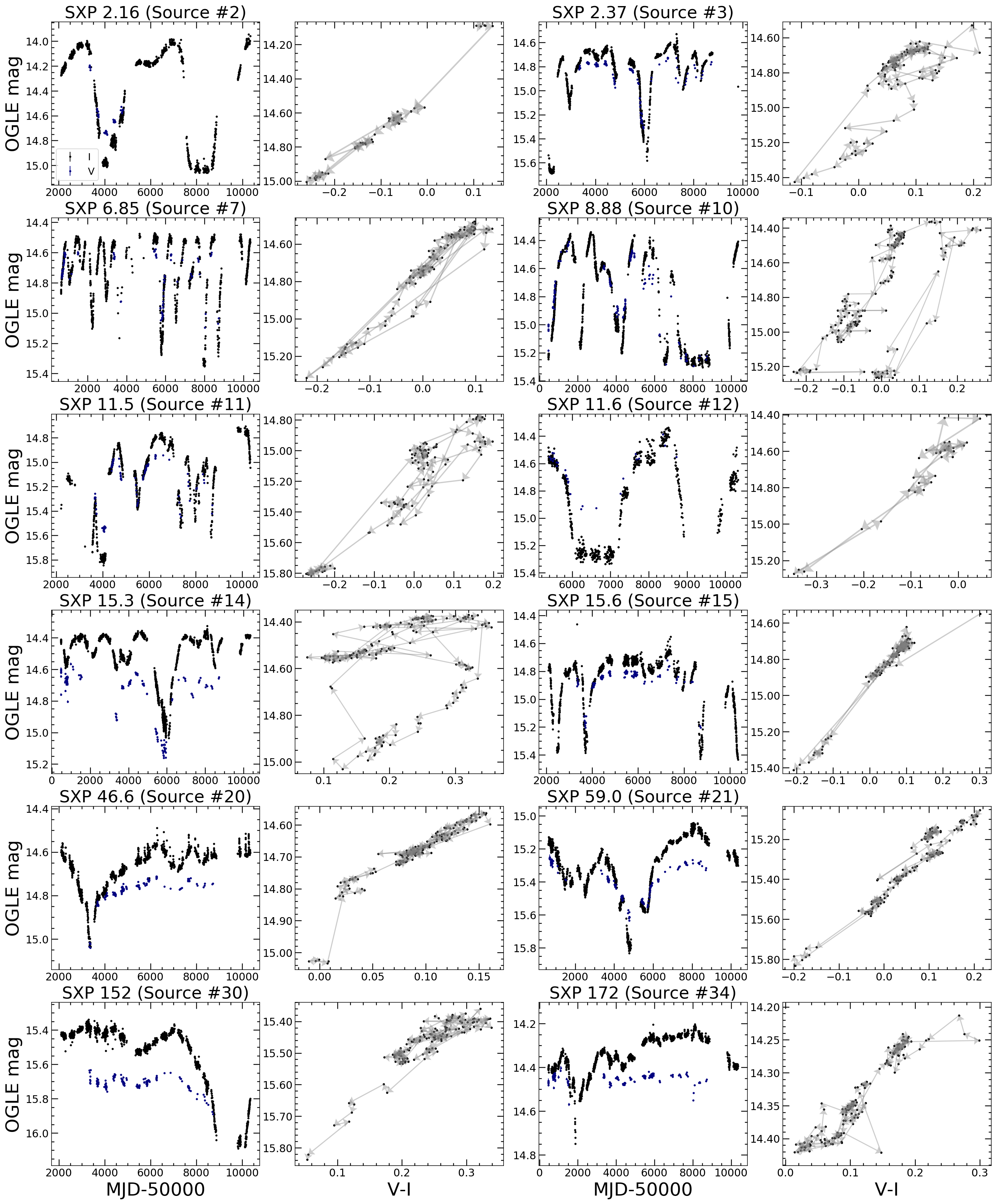}
    \caption{Same as Fig. \ref{fig:type1gallery} but for the first half of the Type 2 sources.}
    \label{fig:type2gallery1}
\end{figure*}
\begin{figure*}
    \centering
    \includegraphics[width=2\columnwidth]{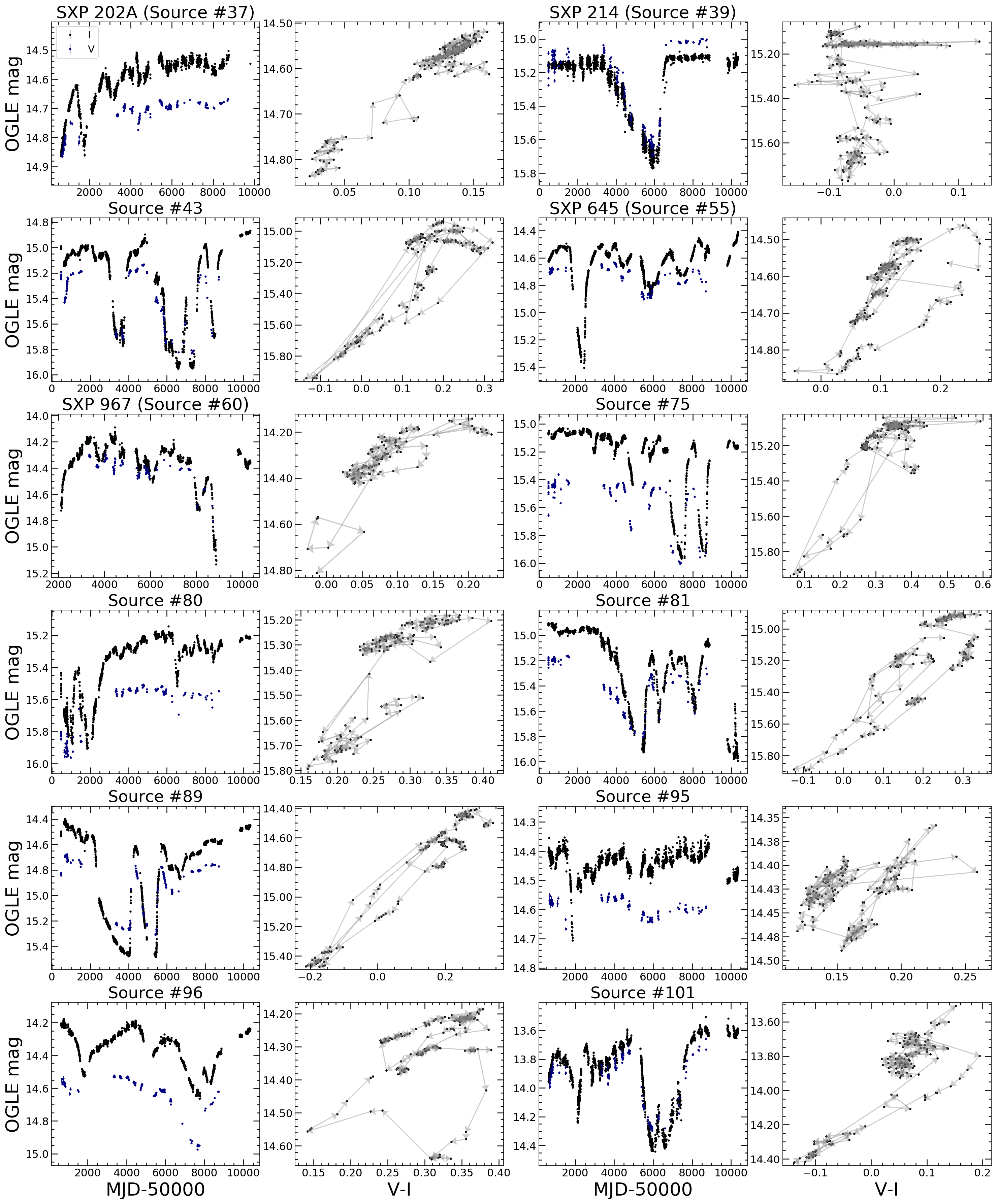}
    \caption{Same as Fig. \ref{fig:type1gallery} but for the second half of the Type 2 sources.}
    \label{fig:type2gallery2}
\end{figure*}
\begin{figure*}
    \centering
    \includegraphics[width=2\columnwidth]{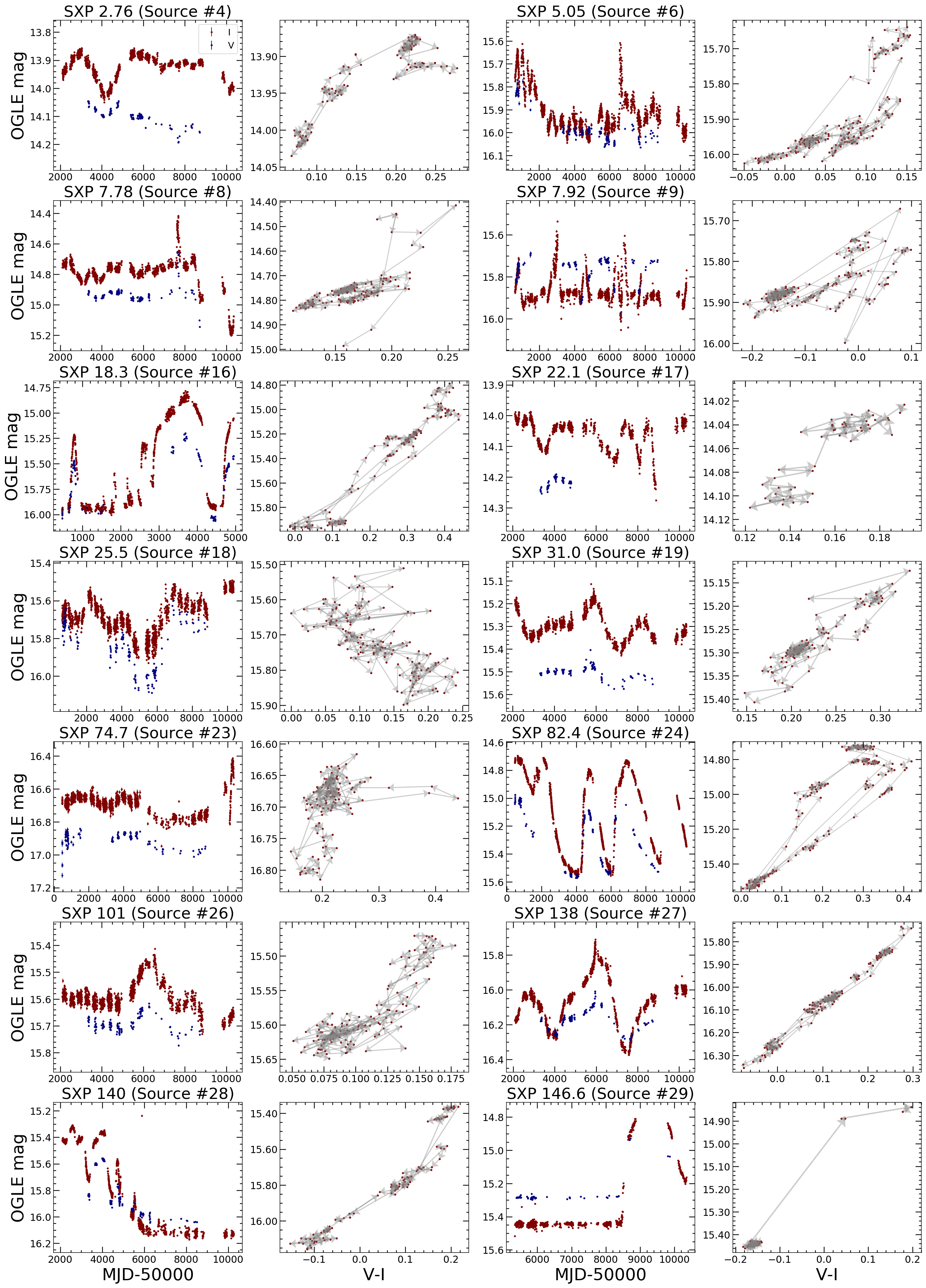}
    \caption{Same as Fig. \ref{fig:type1gallery} but for the first third of Type 3 sources.}
    \label{fig:type3gallery1}
\end{figure*}
\begin{figure*}
    \centering
    \includegraphics[width=2\columnwidth]{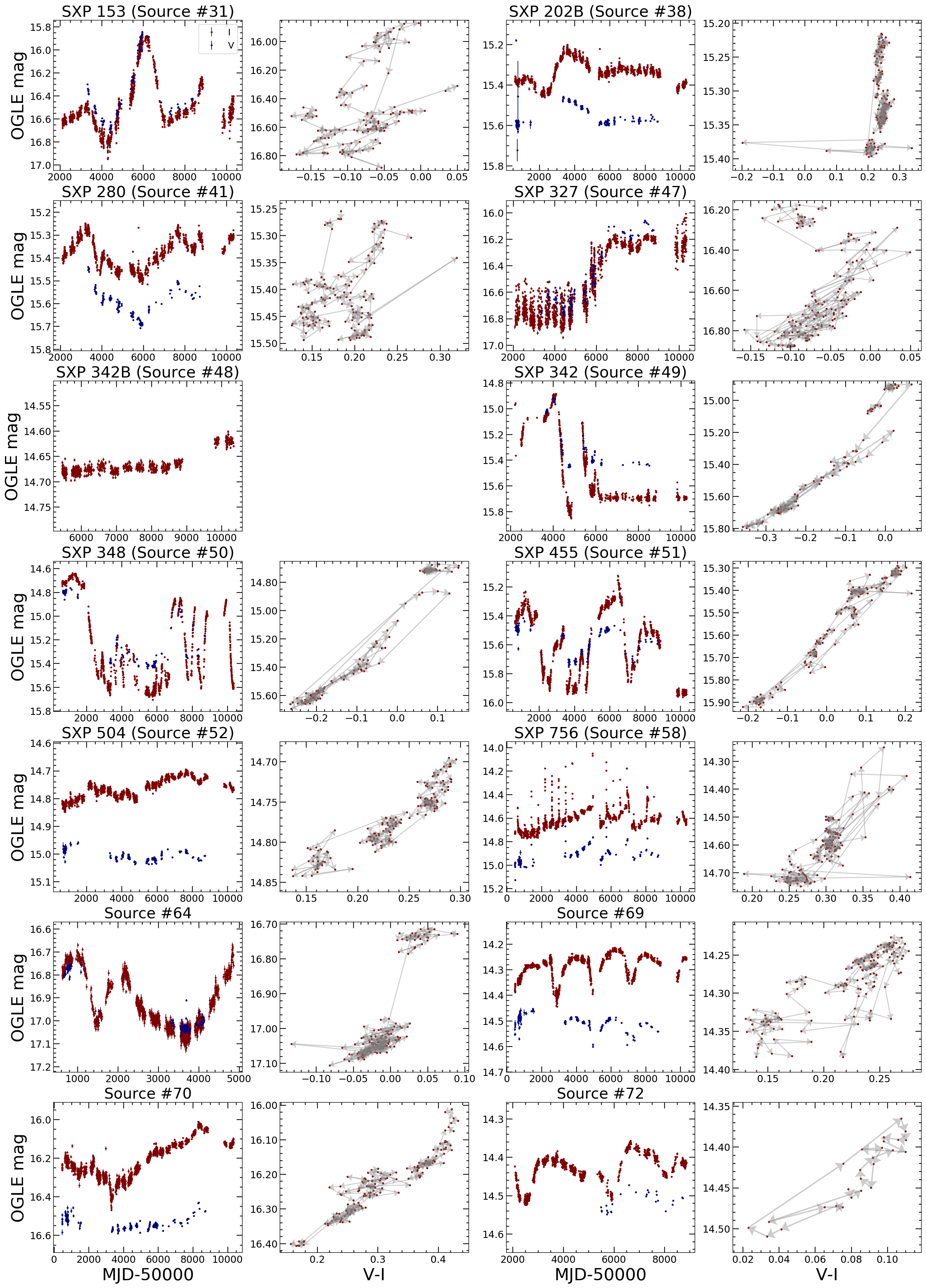}
    \caption{Same as Fig. \ref{fig:type1gallery} but for the second third of Type 3 sources.}
    \label{fig:type3gallery2}
\end{figure*}
\begin{figure*}
    \centering
    \includegraphics[width=2\columnwidth]{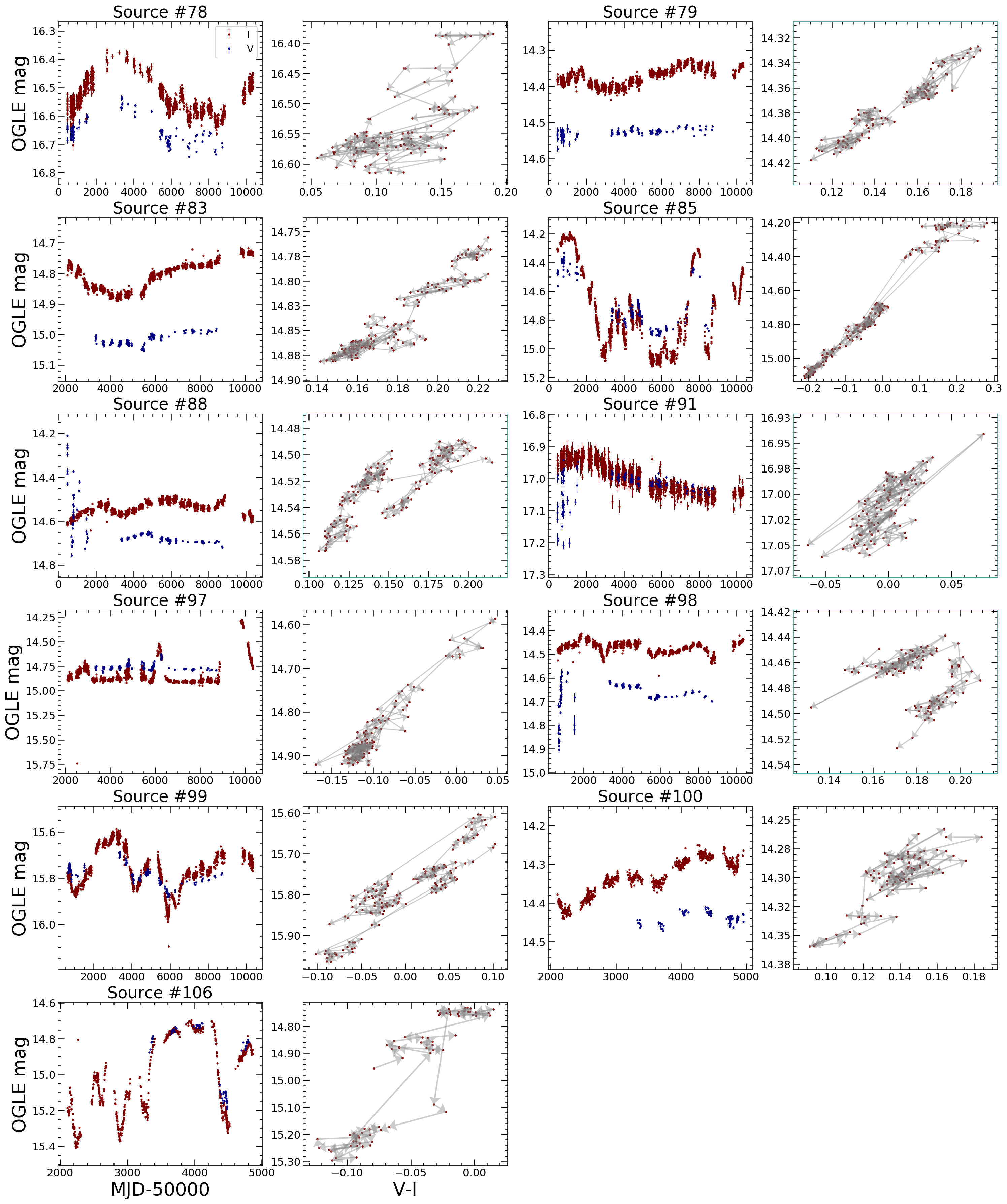}
    \caption{Same as Fig. \ref{fig:type1gallery} but for the final third of Type 3 sources.}
    \label{fig:type3gallery3}
\end{figure*}
\begin{figure*}
    \centering
    \includegraphics[width=2\columnwidth]{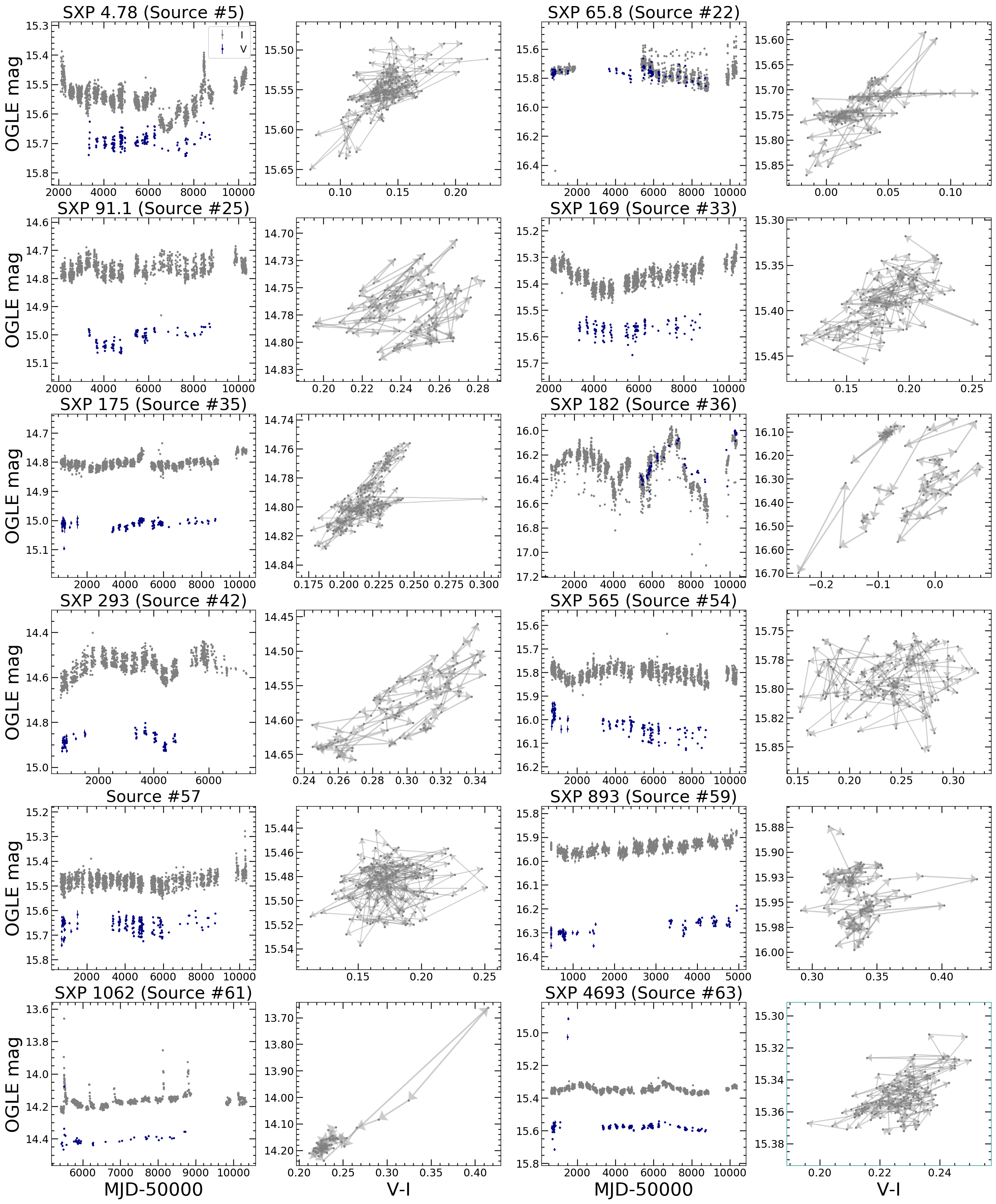}
    \caption{Same as Fig. \ref{fig:type1gallery} but for the first half of Type 4 sources. Color-magnitude diagrams with green outlines exclude OGLE-II data.}
    \label{fig:type5gallery1}
\end{figure*}
\begin{figure*}
    \centering
    \includegraphics[width=2\columnwidth]{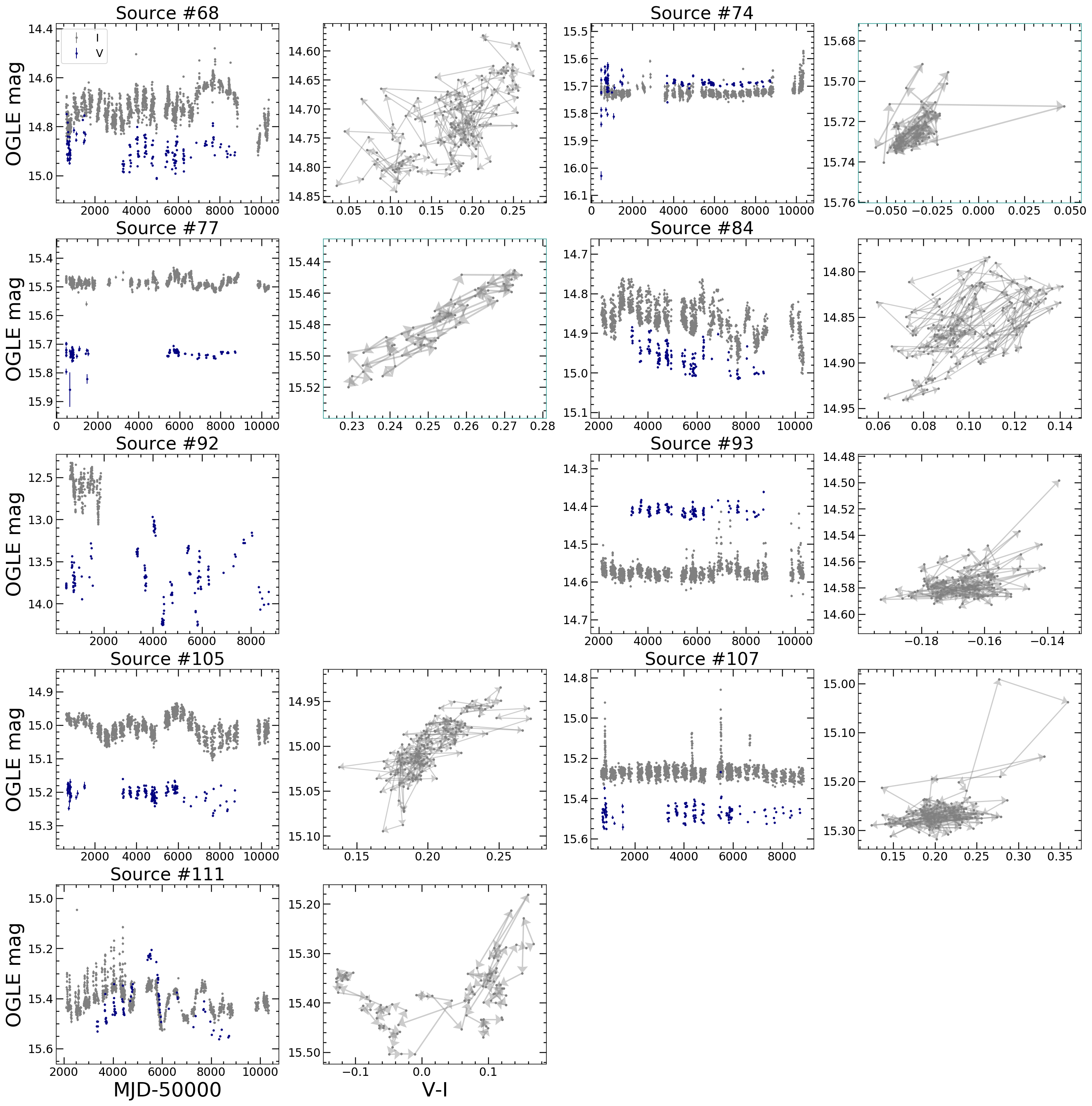}
    \caption{Same as Fig. \ref{fig:type1gallery} but for the second half of Type 4 sources. Color-magnitude diagrams with green outlines exclude OGLE-II data.}
    \label{fig:type5gallery2}
\end{figure*}
\begin{figure*}
    \centering
    \includegraphics[width=2\columnwidth]{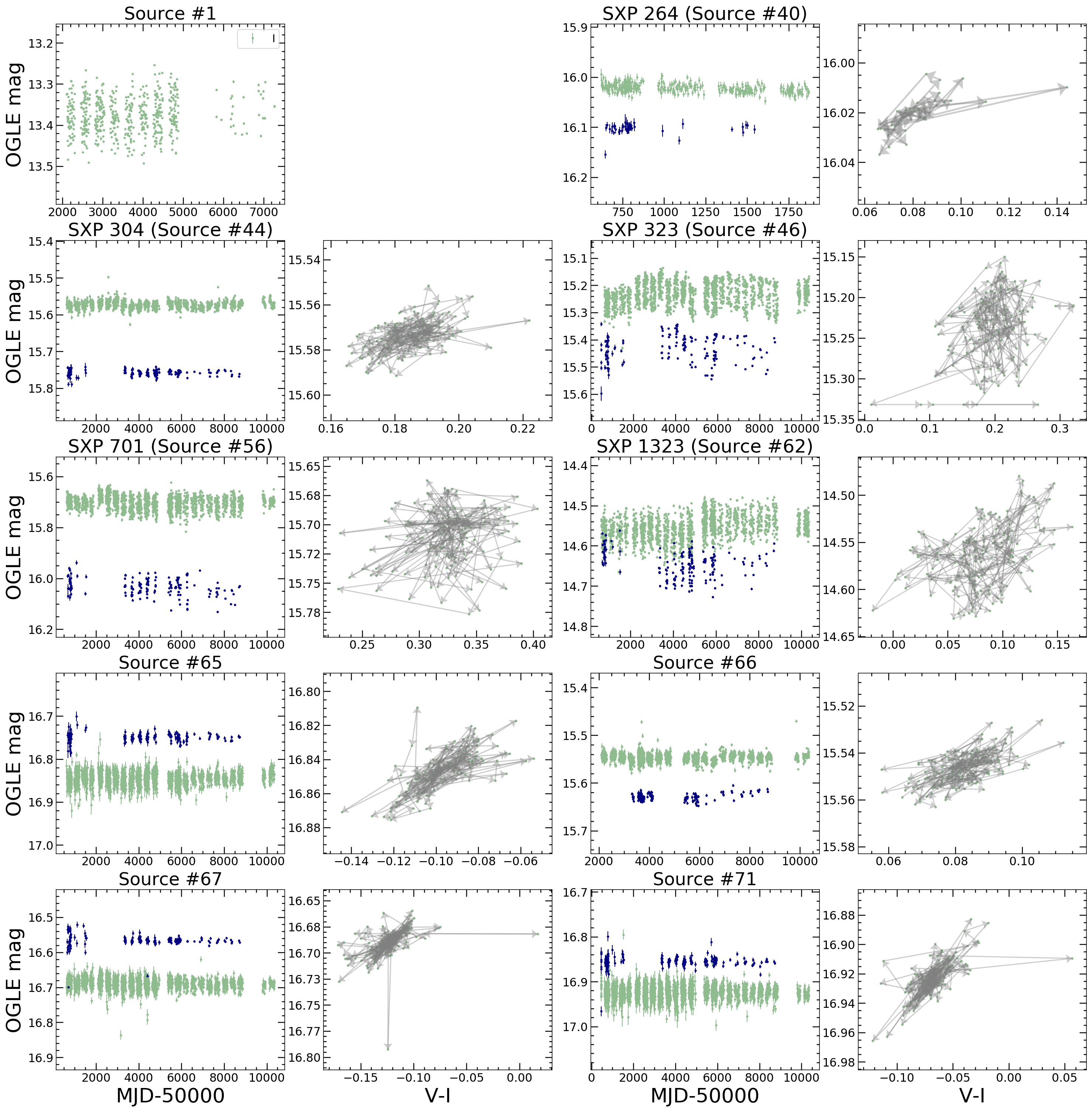}
    \caption{Same as Fig. \ref{fig:type1gallery} but for the first half of the Type 5 sources. Color-magnitude diagrams with green outlines exclude OGLE-II data.}
    \label{fig:type6gallery1}
\end{figure*}
\begin{figure*}
    \centering
    \includegraphics[width=2\columnwidth]{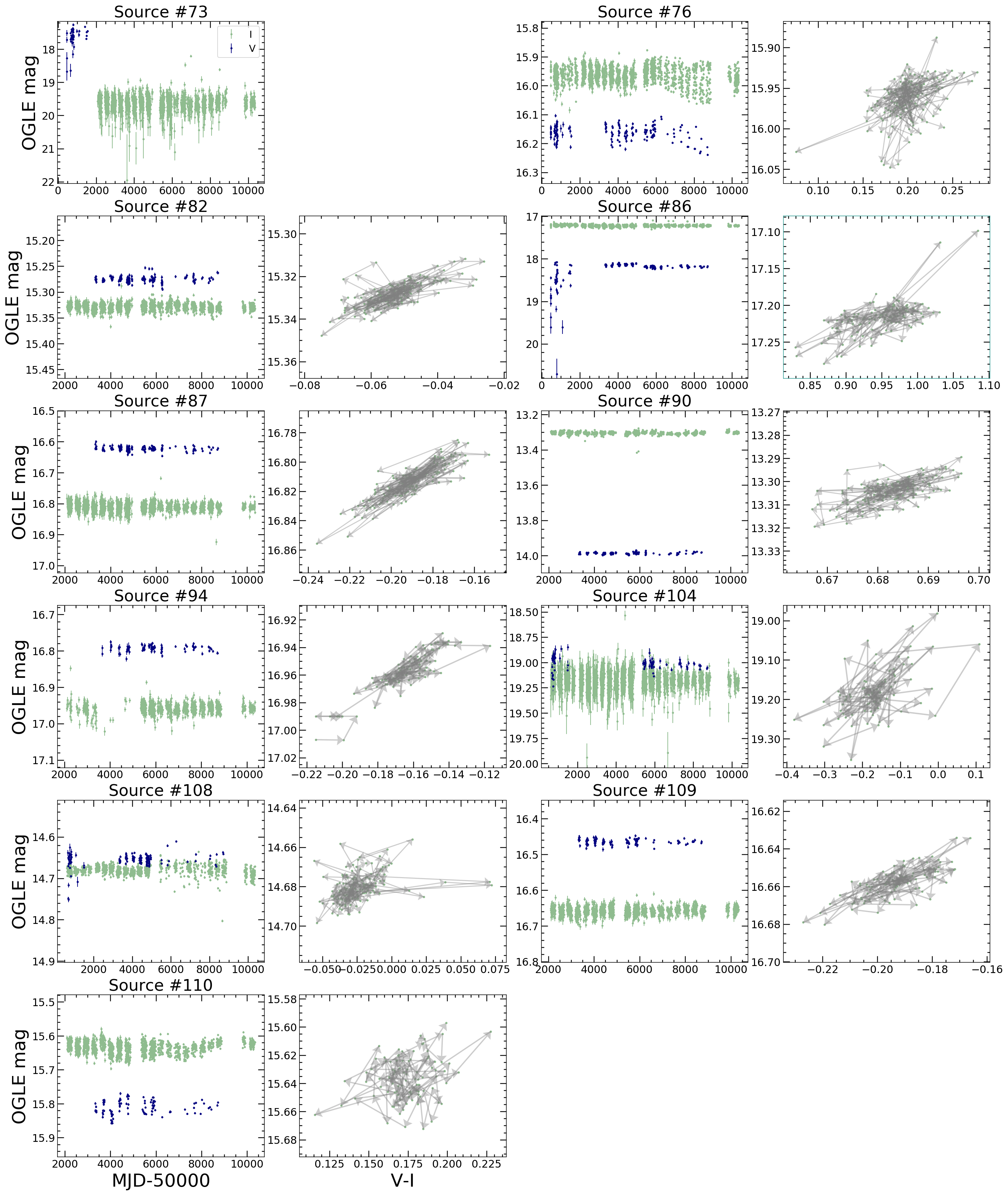}
    \caption{Same as Fig. \ref{fig:type1gallery} but for the second half of the Type 5 sources. Color-magnitude diagrams with green outlines exclude OGLE-II data.}
    \label{fig:type6gallery2}
\end{figure*}

\section{Long data tables} 
\end{document}